\documentclass[structabstract]{aa}  
\usepackage{multirow,bigdelim}
\usepackage{graphicx}
\usepackage{subfigure}
\usepackage{lscape}
\usepackage{longtable}
\usepackage{natbib}
\usepackage{caption}

\bibpunct{(}{)}{;}{a}{}{,}
\usepackage{txfonts}
\begin{document}

\title{$\lambda$=3\,mm line survey of nearby active galaxies}

  \author{R. Aladro
          \inst{1},
S. Mart\'in\inst{2} , D. Riquelme\inst{3}, C. Henkel\inst{3,4}, R. Mauersberger\inst{1},
J. Mart\'in-Pintado\inst{5},  A. Wei{\ss}\inst{3}, C. Lefevre\inst{6},  C. Kramer\inst{7},
M. A. Requena-Torres\inst{3}, R. 
J. Armijos-Abenda{\~n}o\inst{5,8}}

   \institute{European Southern Observatory, Avda. Alonso de C\'ordova 3107, Vitacura, Santiago, Chile.
        \email{raladro@eso.org}
 	\and
        Institut de Radio Astronomie Millim\'{e}trique, 300 rue de la Piscine, Dom. Univ., F-38406, St. Martin d'H\`{e}res, France
       \and
	Max-Planck-Institut f\"{u}r Radioastronomie, Auf dem H\"{u}gel 69, D-53121 Bonn, Germany
	\and
        Astron. Dept., King Abdulaziz University, P.O. Box 80203, Jeddah 21589, Saudi Arabia
        \and
	Centro de Astrobiolog\'ia (CSIC-INTA), Ctra. de Torrej\'on Ajalvir km 4, E-28850 Torrej\'on de Ardoz, Madrid, Spain.
        \and
         LERMA2 \& UMR8112 du CNRS, Observatoire de Paris, 61, Av. de l'Observatoire, 75014 Paris, France
	\and
	Instituto de Radioastronom\'ia Milim\'etrica, Avda. Divina
        Pastora 7, Local 20, E-18012 Granada, Spain.
        \and
        Observatorio Astron\'omico de Quito, Escuela Polit\'ecnica
        Nacional, Av. Gran Colombia S/N, Interior del Parque La
        Alameda, Quito, Ecuador.
}

  \date{Received  04 September 2014 /  Accepted 14 April 2015}

 \abstract
{}
{We aim for a better understanding of the imprints that the
  nuclear activity in galaxies leaves in the
  molecular gas.}
{We used the IRAM 30\,m telescope to observe the frequency
  range $\sim$[86-116]\,GHz  towards the central regions of
  the starburst galaxies M\,83,
 M\,82, and NGC\,253,  the galaxies  
 hosting an active galactic nucleus (AGN) M\,51, NGC\,1068, and NGC\,7469, and the ultra-luminous
 infrared galaxies (ULIRGs) Arp220 and Mrk231. Assuming local thermodynamic equilibrium
  (LTE), we calculated the column densities of 27
  molecules and 10 isotopologues (or their
  upper limits in case of non-detections). }
{Among others, we  report the first tentative detections of  CH$_3$CHO,
HNCO, and NS 
  in M\,82 and, for the first time in the extragalactic medium,
  HC$_5$N in NGC\,253. H$\alpha$ recombination lines were only found in M\,82 and
 NGC\,253. Vibrationally excited lines of HC$_3$N were only detected
 in Arp\,220. CH$_3$CCH emission is only seen in the starburst-dominated galaxies.
By comparison of the fractional
  abundances among the galaxies, we looked for the molecules
  that are best suited to characterise the chemistry of each
  group of galaxies (starbursts, AGNs and ULIRGs), as well as the
  differences among galaxies within the same group.}
{To characterise and compare starburst galaxies, suitable species are
CH$_3$OH and HNCO as tracers of large-scale shocks, dominating early
to intermediate starbursts stages, and CH$_3$CCH, c-C$_3$H$_2$ and HCO as
tracers of UV fields, which control the intermediate to old/post starbursts
phases.
M\,83 shows signs of a shock-dominated environment. NGC\,253 is
characterised by both strong shocks and some UV fields. M\,82 stands out
by bright PDR tracers, pointing to an UV field-dominated environment. 
Regarding AGNs, the abundances of HCN and CN (previously claimed as enhanced
in AGNs) in M\,51 are similar to those in
starburst galaxies, while the HCN/HCO$^+$ ratio is high in M\,51 and
NGC\,1068, but not in NGC\,7469. We neither find a correlation between
the HCN/CS ratio (recently claimed as a possible starburst/AGN
    discriminator) and the AGN activity. However, high enough spatial resolution to
separate their circumnuclear disks (CNDs) from the surrounding
star-forming regions is needed to find molecular abundance trends in AGNs. High
abundances of 
H$^{13}$CN and HC$_3$N, as well as a similarity between the column densities of $^{13}$CO and
C$^{18}$O, are representative of the molecular interstellar medium (ISM)
in the ULIRGs.
Furthermore, the chemistry of Arp\,220 points towards a more
starburst-dominated environment, while that of Mrk\,231 resembles more
the AGNs of our sample.
}

\authorrunning{Aladro et al. (2015)}
\titlerunning{$\lambda$=3\,mm line survey of nearby active galaxies}
  \maketitle{}


\section{Introduction}
\label{sect.Intro}

The creation and destruction of molecular species in the dense
ISM are directly related to a number of physical mechanisms, namely large-scale shocks, cosmic rays,
UV radiation fields and X-rays. Hence, studying molecular abundances
can be a way to infer the importance of these processes for the
    star formation, and to understand how they favour dispersion
        or gravitational collapse
        of compact clouds or their ionisation. In the particular case
of the dense gas found to surround galactic nuclei, one can even
attempt to deduce the nature of their
powering sources  using millimetre(mm)/sub-mm spectra (e.g. \citealt{Martin11}).

Comparing chemical models with observations, it has been claimed that the
abundances of some species, such as HCN and CN, are enhanced in the
presence of X-rays and cosmic ray ionisation fields that are dominant in 
AGNs \citep{Kohno01,Meijerink07,Krips08,Aladro13,Izumi13}. On the other hand, less
energetic radiation from UV fields destroys heavy molecules
like HNCO and CH$_3$OH \citep{Martin06a,Martin08,Martin09}, while at the same time helps creating  others
that are formed through ion-molecule reactions, such as c-C$_3$H$_2$,
CO$^+$ or HOC$^+$ (e.g. \citealt{Savage04,Martin09,Aladro11b}). Shocks
caused by stellar winds, supernovae, or molecular cloud collapse can release
species formed on dust grains to the gaseous phase (e.g. CH$_3$OH,
HNCO, or sulfur-bearing species, \citealt{Viti01,Martin09,Izaskun09}).

Unbiased extragalactic molecular line surveys at mm wavelengths
\citep{Martin06b,Muller11,Aladro11b,Aladro13,Davis13,Watanabe14} allow
us 
to observe many species simultaneously and identify
those that provide the best information about the
physical properties around the nuclear regions of active
galaxies. Some molecules much less abundant 
than the commonly observed species, CO, HCN or HCO$^+$, seem to be
more sensitive to their environments, and show larger
differences in their abundances among galaxy types. They can be even used
 to identify the evolutionary stage of galaxies. For example,
the HNCO abundance  varies by nearly two
orders of magnitude between young and old starbursts, and thus appears
to be an excellent diagnostic tool of their evolutionary state
\citep{Martin09}. CH$_3$CCH, on the other hand, is one of the most abundant species in starburst
galaxies (even more than HCN and HCO$^+$), but has never been
detected in any AGN-dominated source \citep{Aladro11a,Aladro13}.  Yet,
extragalactic line surveys carried out so far have 
focused exclusively on a single type of galaxy, either starbursts
\citep{Martin06b,Aladro11b},  (U)LIRGs \citep{Costagliola11,Martin11,Davis13},
or AGNs \citep{Aladro13}. A
detailed and uniform comparison 
of different types of active galaxies in this context is still
missing. Here, we therefore aim at a
comparison of the molecular ISM in a sample of
galaxies that includes starbursts, AGNs and ULIRGs.

In this paper we present the first systematic unbiased spectral line
survey of a sample of galaxies selected with respect to
their  nuclear activity. It is organised as follows: In Sect.~\ref{galaxysample} we
introduce our galaxy sample. Section~\ref{sect.Obs} describes the
observations and data reduction. Section~\ref{analysis} details the data analysis,
including a discussion about the molecular source sizes of the
galaxies (Sect.~\ref{sourcesizes}), the line identification and line profiles
(Sect.~\ref{lines}),  spectral model fits, and rotational temperatures
(Sect.~\ref{madcuba}). In
Sect.~\ref{results} we summarise the detections
(Sect.~\ref{detections}), and present the column densities
calculated under the LTE assumption
(Sect.~\ref{columndensities}). We indicate the potential key species
for further extragalactic studies in
Sect.~\ref{keyspecies}. In Sect.~\ref{discussion}
we  detail our findings for  starburst
galaxies (Sect.~\ref{SB}), AGNs (Sect.~\ref{AGNs}), and ULIRGs
(Sect.~\ref{ULIRGs}). Finally, we summarise our main conclusions in Sect.~\ref{conclusions}.

\section{The galaxy sample}
\label{galaxysample}

We selected a sample of eight galaxies based on their luminosity and galactic nuclear
activity, following these characteristics: (a) the
galaxies are in the local universe ($z<0.05$,
D $<$ 200\,Mpc), (b) they show
bright molecular emission ($T_{\rm
  MB}^{CO(1-0)} \textgreater 50$ mK) and cover a wide range of infrared luminosity
($L_{\rm IR}=[10^9-10^{12}]\,L_\odot$), thus including low luminosity
galaxies, LIRGs and ULIRGs; (c) the sample encompasses
the two main types of galactic nuclear activity: AGNs and starbursts; 
  (d) it includes starbursts and ULIRGs in
different stages of  evolution; (e) it includes AGNs of both Seyfert
(Sy) 1 and Sy\,2 types. The eight galaxies are well-known extragalactic
objects studied across almost the whole electromagnetic spectrum,
and in particular at mm/sub-mm
wavelengths. Moreover, these galaxies are often considered as
archetypes in the local universe of the different nuclear
activities/stages of evolution mentioned above.  Some of their main
properties are shown in Table~\ref{galaxies}.

Our sample includes the three starburst galaxies  M\,83, NGC\,253, and M\,82.
The last two were targets of line surveys at $\lambda=2$ and 1.3 mm done with the
IRAM 30\,m telescope by \citet{Martin06b} and \citet{Aladro11b}. These
two studies confirmed that the molecular ISM in the north-eastern
molecular lobe of M\,82 shows typical
features of an evolved starburst, where the gas is heavily pervaded by
strong UV fields, while the nuclear region of NGC\,253 is more
influenced by low-velocity shocks.  NGC\,253 hosts notable large scale
outflows \citep{Turner85,Bolatto13} which indicate that its starburst
is not in its initial state, though not as evolved as the M\,82 one.
With the present $\lambda$=3\,mm
observations, these two galaxies
are the best studied at millimetre wavelengths so far.
 We also observed M\,83 as an example of a young
 starburst \citep{Martin09}.
It has a lower
star formation rate (2.5\,$\rm M_\odot$\,yr$^{-1}$, \citealt{Walter08}) than
NGC\,253 and M\,82, and a
star formation episode lasting (only) around 6 Myrs \citep{Houghton08}. Thus its
chemical composition is expected to be somehow different from those of
NGC\,253
and M\,82. With these three starburst galaxies, we aim to 
characterise better the changes of the molecular gas composition as the starburst
processes evolve. 

As galaxies containing AGNs, we include NGC\,7469 (Sy\,1), NGC\,1068 (Sy\,2), and M\,51
(Sy\,2).  The results of the NGC\,1068 survey were
already published in \citet{Aladro13}, and we refer the reader to that
paper for a discussion of its isotopic
line ratios and a chemical modelling of its central gas. 
Both NGC\,7469 and NGC\,1068 have a starburst ring at $\sim$1\,kpc from the
central AGN, and have very similar observed properties (see
e.g. Table\,2 of \citealt{Wilson91}). On the other hand, M\,51 shows
several star-forming regions along
its spiral arms \citep{Bastian05}. In the three cases, the angular
resolution of the 30\,m beam ($\sim25''$) did not allow us to
separate the emission coming from the central AGNs and those
of the surrounding starbursts. Thus, our data might be showing not
only the molecular composition of their AGN but also some contribution
from the starburst regions. This might be particularly significative
in the case of NGC\,7469 due to its farther distance, although
estimations of the emission coming from starburst regions at our
frequencies are only available for NGC\,1068 (see details in Sect.~\ref{AGNs}).

To complement our sample, we observed two of the closest ULIRGs,
Arp\,220 and Mrk\,231. Vigorous star forming regions and black
 holes coexist in the centres of ULIRGs. Still, the main powering source of their nuclei remains unclear in
 most cases. Arp\,220 is in an early stage
of the merging process, with a double nucleus  that is not resolved by our
observations. An interferometric line survey done towards this galaxy by \citet{Martin11}
pointed to starburst processes as the more likely heating mechanism
of its two centres (see also \citealt{Greve09}). In contrast, Mrk\,231
is in a later stage of merging, where the two nuclei of the individual galaxies
are already fused, and molecular-rich large-scale
outflows and super-winds are clearly present \citep{Aalto12}. AGN activity likely
prevails in
this galaxy (e.g. \citealt{Gallagher02,Van10}),
and its nucleus seems to host a Sy\,1 object \citep{Boksenberg77}.

\section{Observations and data reduction}
\label{sect.Obs}

\begin{figure*}
\centering
\includegraphics[width=0.7\textwidth,angle=0]{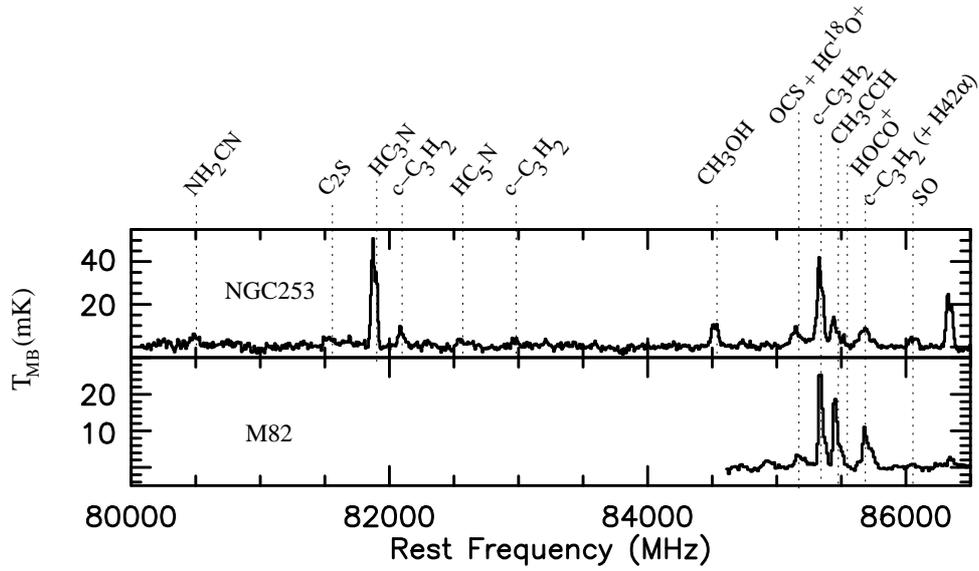}
\caption{Lowest frequency range observed exclusively towards NGC\,253 and M\,82 (the rest
  of the galaxies were observed from $\sim86$\,GHz).}
\label{lowfreq}
\end{figure*}

The observations were carried out with the IRAM 30\,m
telescope\footnote{IRAM is supported by INSU/CNRS (France), MPG
  (Germany), and IGN (Spain).} (Pico Veleta Observatory, Spain)
between June 2009 and March 2012. We observed the eight
galaxies in the frequency range $\sim$[86-116]\,GHz. For NGC\,253 and
M\,82 we also included lower frequencies from ~80\,GHz and ~85\,GHz, respectively. The reduced spectra are
shown in Figs.~\ref {AllSurveys1}  - ~\ref {HC5N}. The
half-power beam width (HPBW) corresponding to the survey frequencies ranged
from 29$''$ to 21$''$. We used the band E0 of the EMIR receiver \citep{Carter12} and
the WILMA autocorrelator. This receiver-backend configuration allowed
us to cover 8\,GHz simultaneously in the two orthogonal linear 
polarisations, and led to an original channel-width spacing of
$5-7$\,km\,s$^{-1}$. The observed $\sim$30\,GHz were covered by
    5 tunings with a minimum overlap of
    0.8\,GHz.  For some galaxies,  an extra tuning has been used to increase the
    signal to noise in parts of the spectrum showing faint
    lines. The observations were done by wobbling the secondary mirror with a switching frequency of 0.5\,Hz and a beam
throw of 110$''$ in azimuth (80$''$ in the case of Arp\,220).

 The weather was variable during the different observing runs, from
 excellent (pwv$\sim$1\,mm) to  mediocre (pwv$\sim$6\,mm) weather conditions.  We checked the pointing accuracy every hour
 towards several nearby bright continuum sources. The pointing
 corrections were always $<$5$''$. The focus was also checked and
 corrected at the beginning of each run and during sunsets and sunrises.

\begin{table*}
\caption{Main properties of the galaxies}
\centering
\begin{tabular}[!h]{lcccccccccccc} 
\hline
\hline
Galaxy & $\alpha$ (J\,2000)           &  $\delta$ (J\,2000)     & D$^
a$ & $V_{\rm LSR}^ b$     &$\theta_{\rm s} ^ c$ & SFR$ ^ d$  &
$L_{\rm IR} ^e$ & Activity type\\
            & h:m:s   &  $^{\rm o}$:$'$:$''$ &  Mpc    &   km\,s$^{-1}$ &
            $ ''$ /pc  & M$_\odot$\,yr$^{-1}$  &     
$L_\odot$ &\\
\hline
M\,83   & 13:37:00.93  & -29:51:56.40 & 4.5&513&15 / 327&2.5&
{\bf{$1.2\times10^{10}$}} & Starburst\\
NGC\,253         & 00:47:33.12  & -25:17:18.60 & 3.9 & 250
&20 / 378 &3.6 &
$2.8\times10^{10}$ & Starburst\\
M\,82-NE       & 09:55:51.90  & 69:40:47.00  & 3.5 & 300 &12 /204 &10 & 
{\bf{$5.9\times10^{10}$}}& Starburst \\
M\,51            & 13:29:52.70  & 47:11:43.00  & 8.4 & 470&15 / 611& 2.5  &
{\bf{$2.6\times10^{10}$}}& AGN \\
NGC\,1068        & 02:42:40.90  &-00:00:46.00  & 14.4 & 1100
&4 / 279&0.4 &
$1.9\times10^{11}$ & AGN + Starburst \\
NGC\,7469          & 23:03:15.60  & 08:52:26.00  & 29.5 & 4892&6 / 858&30&
$3.9\times10^{11}$ & AGN + Starburst \\
Arp\,220         & 15:34:57.08  & 23:30:11.30  &70.0 & 5350 &2 / 679&240 &
$1.6\times10^{12}$ & Starburst\\
Mrk\,231         & 12:56:14.20 & 56:52:25.00& 170.0 &12173 &2 / 1640& 220 &
$3.2\times10^{12}$ & AGN \\

\hline
\end{tabular}

\begin{list}{}{}
\item For M\,82 we observed the northeastern (NE) molecular
  lobe at an offset position ($\Delta \alpha, \Delta \delta) =
  (+13.0'', +7.5''$) with respect to its dynamical centre. $^a$
  Distances taken from the NASA/IPAC Extragalactic Database; $^b$
  Systemic Local Standard of Rest (LSR) radial velocities following the optical convention, taken from SIMBAD. 
$^c$ Source sizes and their corresponding spatial scales for the
assumed distances. $^d$ Star
  formation rate obtained from \citet{Walter08} for M\,83;
  \citet{Strickland04} for  NGC\,253 and M\,82; \citet{Schuster07} for
  M\,51; \citet{Esquej14} for NGC\,1068;  \citet{Genzel95} for 
  NGC\,7469; \citet{Anantharamaiah00} for Arp\,220; \citet{Taylor99} for Mrk\,231. We note that this parameter
  should only give a rough idea of the activity, as it was calculated for
  different volumes in each galaxy. $^e$ Infrared luminosities taken
  from \citet{Sanders03}.
\end{list}{}{}

\label{galaxies}
\end{table*}

The data were first calibrated to the
antenna temperature ($T_{\rm A}^*$) scale using the chopper-wheel
method \citep{Penzias73}. The observed spectra were then converted to main beam
temperatures ($T_{\rm MB}$) using the relation $T_{\rm MB}=(F_{\rm
  eff}/B_{\rm eff})\,T_{\rm A}^*$, where $F_{\rm eff}$ is the forward
efficiency of the telescope, whose values were between 0.94 and 0.95,
and $B_{\rm eff}$ is the main beam efficiency, ranging from 0.77 to
0.81
\footnote{http://www.iram.es/IRAMES/mainWiki/Iram30mEfficiencies}. 

Each individual spectrum was analised separately. After
    eliminating  bad
    channels or spectra containing strong ripples,
    all the spectra tuned at the same frequency were averaged and 
baselines of orders between zero and two were subtracted 
to each chunk of 8\,GHz. We note however that in the case of
    M\,82 the baseline correction was not good enough to remove some small
    ripples in the range $\sim[107-109]$\,GHz. Higher order baselines did not
    improve the resulting spectra. Fortunately, comparing with
    the rest of the galaxies, no important lines are expected in that range.

Due to the broad line-widths of these galaxies
($\ge$100\,km\,s$^{-1}$), the final spectra of M\,51, M\,82, M\,83,
NGC\,1068 and NGC\,253 were smoothed to $11-14$\,km\,s$^{-1}$, 
those of Arp\,220 and NGC\,7469 were smoothed to
$22-27$\,km\,s$^{-1}$, and the Mrk\,231 data to $\sim$60\,km\,s$^{-1}$.
The rms achieved
for all galaxies was in average  $1-2$\,mK ($\le 10$\,mJy\,beam$^{-1}$) across the whole
survey at the final velocity resolutions. The data were also
corrected to first order for beam dilution effects as 
\begin{equation}
T_{\rm B}=[(\theta^2_{\rm s}\,+\,\theta^2_{\,\rm
  b})\,/\,\theta^2_{\,\rm s}]\,T_{\rm MB}\,,
\label{eq1}
\end{equation}
where $T_{\rm B}$ is the source averaged
brightness temperature, $\theta_{\,\rm s}$ is the molecular source size
(see Table ~\ref{galaxies} and Sect.~\ref{sourcesizes} for details),
and $\theta_{\,\rm b}$ is the beam size. Gaussian distribution of
the emission was assumed. We did not correct the data for the gain-elevation
curve and for the gain losses due to a large wobbler throw because
these corrections are only minor at 3\,mm wavelength.

\section{Data analysis}
\label{analysis}

\subsection{Source sizes}
\label{sourcesizes}
The source size of the molecular emission in each galaxy can be estimated in
two ways. One
is through high angular resolution interferometric maps of the region
of all the species, and the
other is by comparison of the observations carried out with two (or more)
telescopes having different beam sizes at the same frequency (so
$\theta_{\rm s}$ can be estimated from Eq.~\ref{eq1}).
Unfortunately, the interferometric
maps performed so far for our sample of galaxies encompass a few
abundant species only, such as CO, $^{13}$CO, or HCN. Besides,
most of the line transitions studied here were previously
observed (if detected at all) with only one single-dish
telescope. 

To simplify the situation, we adopt the same source size for all the molecules
detected in a galaxy. We estimated their $\theta_{\rm s}$ from 
interferometric maps of
$^{12}$CO\,$(1-0)$, HCN\,$(1-0)$,  HCO$^+$\,$(1-0)$, $^{13}$CO\,$(1-0)$, and
HCO\,$(1_{01}-0_{00})$ 
(\citealt{Downes98} for Arp\,220 and Mrk\,231,
  \citealt{Kohno96} for M\,51, \citealt{Burillo02} for the NE lobe of M\,82, \citealt{Muraoka09} for
  M\,83, \citealt{Helfer95}, \citealt{Schinnerer00} for NGC\,1068,
  \citealt{Knudsen07} for NGC\,253, and \citealt{Davies04} for NGC\,7469). 
The values are listed in Table~\ref{galaxies}. 

The  source size has a direct impact on the resulting brightness
temperature (Eq.~\ref{eq1}), and consequently on the source averaged column
density.The larger (smaller) the source size that is assumed, the
lower (higher) the column densities that are derived.
To work around this, we base our discussion on column density ratios
between species. Our estimates of these ratios would be only sensitive to differences in the emission extent between species, on which there is not enough information.

\subsection{Line identification, line profiles, and Gaussian fits}
\label{lines}
We used the MADCUBA\_IJ
software\footnote{http://cab.inta-csic.es/madcuba}
which includes the CDMS and JPL catalogues \citep{Muller01,Muller05,Pickett98} 
to identify the lines. The procedure was
similar to what is described in detail in \citet{Aladro11b}, and also
shown in the NGC\,1068 survey \citep{Aladro13}.

We used CLASS\footnote{http://www.iram.fr/IRAMFR/GILDAS/} to check whether
there were lines coming from the image band of the spectra. The
image band rejection of the E0 receiver is $>$10\,dB, which
ensures that most of the lines from the image band, except for the strongest ones, are not
entering the signal band. We found
that only the spectra from NGC\,253, M\,83 and M\,82 are affected by
features coming from the image
band. Few of them were blended with other
transitions from the signal band. 
For those, whenever the separation was large enough, we fixed the
observed line parameters in the signal band to estimate their
contribution to the image band line. 
Details of the fitting parameters are shown in 
Appendix~\ref{LongTables}.

Among the eight galaxies, only M\,83 and Mrk\,231 show simple Gaussian line profiles.
The bright lines in Arp\,220 (e.g. HCN, HCO$^+$, CS) have a clear
    double-peak profile. It is not clear whether this might be
        due to the unresolved double nucleus or to
        absorption effects, as
        reported in previous interferometric studies \citep{Sakamoto09,Tunnard15}. Fainter
    lines also show a deviation from Gaussian profiles, though not so
    clear as those which are optically thicker. Mrk\,231
    is also expected to be affected by absorption, but its lines are
    comparatively much fainter, and thus this effect, if present, is not so
    evident. Regarding NGC\,253, the observed circumnuclear region has five clumps of warm dense gas, but
interferometric  observations are needed to resolve them
(e.g. \citealt{Sakamoto11}). With the 30\,m telescope, the integrated spectrum
of NGC\,253 appears as two
blended Gaussians. On the other hand, the `two-horned' line profiles of
M\,51 and NGC\,7469 include not only the emission from the very centre, but possibly also
that coming from the inner arms and the surrounding molecular
disks  \citep{Matsushita98, Davies04}. 
NGC\,1068 has a similar physical structure as NGC\,7469 (a
circumnuclear disk, surrounded by a starburst ring at $\sim$1kpc from
the centre). In addition, the gas
dynamics in the CND of NGC\,1068 are quite complex, and some of the
brightest lines, as seen by high-resolution observations, need several (two to four) Gaussian components to
better fit the profiles \citep{Krips11,Burillo14}. Finally,
while our M\,82
observations are centred in the North-Eastern (NE)
molecular lobe of the nucleus, some emission from the
very centre enters the 30\,m beam. This
translates into a bump at low velocities of some of the lines that we can
easily separate from the NE emission (see
\citealt{Aladro11b} for details). To
homogenize the data, we fitted a single Gaussian to all lines detected
in the survey. This simplification has little impact on the
derived column densities of the species, as shown by \citet{Martin11}
and \citet {Aladro11a} for Arp\,220 and NGC\,253 respectively.
However, we made sure that the results obtained by fitting one or more Gaussians to
the lines give differences in the overall column densities that
are less than 10\% of their values.

\subsection{Spectral line model fit}
\label{madcuba}
Molecular emission was modelled using the SLIM package within MADCUBA\_IJ (Mart{\'i}n et al. in prep.).
The fit is performed in the parameter space of column density, rotational
temperature, velocity and width of the line to the actual
emission. Thus integrated and peak line intensity, as well as opacity, are derived from
those parameters. The modelling assumes LTE, but not optically thin
emission. However, the fixed  values of the source
    sizes and rotational temperatures make the fitted lines optically
    thin and the line profiles Gaussian.
A Gaussian fit has
been performed to the CO\,$(1-0)$ lines, but as it is highly affected by
opacity, which impact the column density estimate, we do not include CO in our
results and discussion. For the rest of the species, we fitted the molecules one by one,
including in each case all the transitions along the observed
frequency range (in case there are more than one). As an exception,  
blended features were fitted simultaneously in order to take
into account the contribution of each one to the total integrated areas.

The rotational temperature ($T_{\rm rot}$) of a given species can only be constrained
if there are observations of several rotational transitions. Additionally, if
the molecule is faint, the error
associated to its calculated $T_{\rm rot}$  could be, by far, larger than the
value itself. As most of the species have only one
transition in our survey, we fixed $T_{\rm rot}$=10\,K. On average, as
seen in our previous works,
the rotational temperatures of the detected species
 are not expected to be much lower
or higher than this value, though some exceptions are observed, such
as seen for NH$_2$CN and CH$_3$CCH in NGC\,253 and M\,82
\citep{Martin06b,Aladro11b}, HC$_3$N and c-C$_3$H$_2$ in Arp\,220
\citep{Martin11}, and CN in Mrk\,231 and Arp\,220 \citep{Henkel14}. The uncertainty added to the calculated column
    densities by assuming a fixed rotational temperature is not
    critical when comparing fractional abundances among galaxies;
    multi-line observations of many molecules in the nuclei of
    different galaxies and in the GC show that species have also
    very similar excitation conditions in different sources
    (e.g. \citealt{Aladro11b,Armijos14}). Thus the associated
    rotational temperature of a given molecule is expected (or
    measured) to be similar for all the galaxies of our
    sample, being the variations of the $T_{\rm rot}$ proportional to the differences
        in the gas temperature among sources. For this reason, comparing
        fractional abundances instead of column densities, makes the
        analysis less sensitive to these changes.


Apart from the species mentioned before, there is the special case of the vibrationally excited lines of
HC$_3$N, detected (only) in Arp\,220. They are expected to trace gas with
temperatures above 100\,K \citep{Costagliola10,Martin11}. For those vibrational
transitions, we did calculate the temperature in a similar
  way as for the rotational temperatures, and obtained
$T_{\rm vib}=190\pm20$\,K.

Warm kinetic temperatures ($T_{\rm kin}$) estimated in the nuclear regions of these galaxies, such
as 120\,K for NGC\,253 \citep{Bradford03} or
$\ge$150\,K for Mrk\,231 \citep{Van10}, come
from higher frequency observations and are representative of a
small fraction of the gas only. Therefore, our assumption is still valid
for the bulk of the molecular gas. Note, in any case,
that rotational temperatures are lower than $T_{\rm kin}$, as the H$_2$
densities are below the critical densities.
Also, previous extragalactic surveys used similar
assumptions, such as $T_{\rm rot}$=12$\pm$6\,K for NGC\,253 \citep{Martin06b},
20$\pm$10\,K for M\,82 \citep{Aladro11b}, and 10$\pm$5\,K for NGC\,1068
\citep{Aladro13}.

\section{Results}
\label{results}
\subsection{Summary of detections }
\label{detections}

\begin{figure*}
\centering
\includegraphics[width=\textwidth,angle=0]{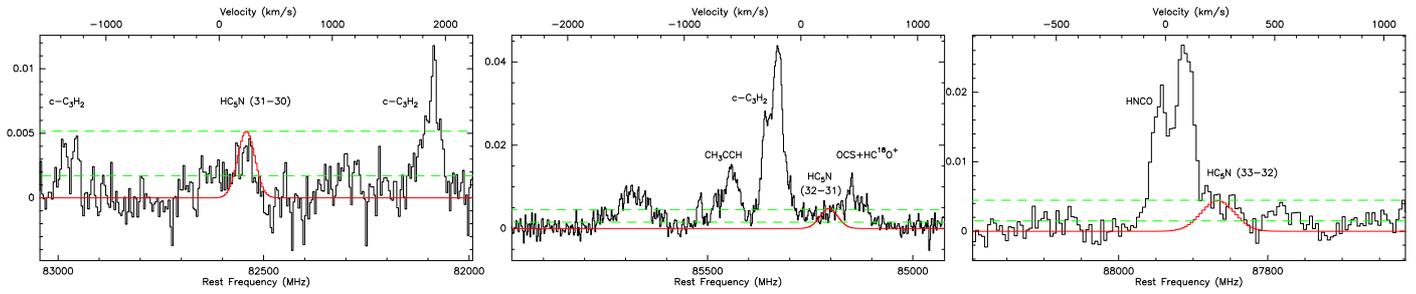}
\caption {Zoomed spectra of NGC\,253 showing the Gaussian fits to the three strongest
  transitions of HC$_5$N detected tentatively. Dashed lines are $1\sigma$ and
  $3\sigma$ noise levels.}
\label{HC5N}
\end{figure*}

In agreement with our previous surveys
    \citep{Aladro11b,Aladro13}, we consider ``detected'' lines as those with a signal to noise
    ratio SNR\,$>$\,3\, and whose profiles can be clearly 
    fitted by a Gaussian. Lines are considered ``tentatively detected''
    when having 1\,$<{\rm SNR}\le3$, but can also be fitted with Gaussian
    profiles, whose velocities and line widths are in good agreement
    with the ones obtained for detected lines (e.g. HC$_5$N lines, see
     Fig.~\ref{HC5N}). ``Non-detected'' lines are
    those with SNR\,$\le$\,1\, for which it is not
    possible to fit reliable Gaussian profiles, as the lines are
    dominated by the spectral noise.

Thirty-seven species (including isotopologues) were identified in NGC\,253, which make it the
most prolific galaxy of our sample in terms of molecular
complexity. We detected the $^{13}$C bearing isotopologues of CO, HCN, HCO$^+$,
HNC, CN, and CS, the $^{18}$O bearing isotopologues
of CO and HCO$^+$, the $^{17}$O isotopologue
of CO, and the $^{34}$S bearing isotopologue of CS. Furthermore,
$^{13}$CN, SO$_2$, NH$_2$CN, HOCO$^+$, HC$_5$N, H$_2$CS, and C$_2$S
were only detected in NGC\,253. Tables~\ref{Nmol1} and ~\ref{Nmol2}  list the total column densities of the thirty-six
molecules and isotopologues 
in the eight galaxies (excluding CO, see Sect.~\ref{lines}).
Molecules detected (tentatively or not) for the first time in each of the galaxies are marked
in boldface.

Although tentative, this is
the first time that HC$_5$N is detected outside the Milky Way. All the
seven brighter 
transitions based on our synthetic spectra of HC$_5$N appear with
measured intensities at 
$1-3~\sigma$ level (see Table~\ref{TablaNGC253}, and Fig.~\ref{HC5N}). Deeper observations
would be needed to confirm this. We note, however, some
    inconsistencies in the identification of this species
    between \citet{Meier15} and this work. The features that we
    identify as the HC$_5$N  transitions $(J-J')$=$(33-31)$,
    $(37-36)$, $(38-37)$ at 87.8, 98.5, and 101.2\,GHz respectively, are
  tentatively identified by \citet{Meier15} as NH$_2$CHO, CH$_3$CH$_2$CN, and
  CH$_3$SH, also in the centre of NGC\,253. We do not detect emission of those molecules in our
  survey, although we cannot rule out that this might be due to differences in
                    sensitivity between the IRAM 30\,m telescope and ALMA. However,  we cover a broader range of frequencies where other
HC$_5$N lines are also consistently detected, which makes us favour our identification. HC$_5$N was previously observed towards
Galactic dense cores
of dark clouds, and seems to trace the initial
conditions of star formation (e.g. \citealt {Suzuki92,Rathborne08}).

In addition, we found that c-C$_3$H$_2$, NS and, especially, CH$_3$CHO were 
detected in NGC\,253 and M\,82, but not in the rest of the galaxies of our
sample. Outside the Milky Way, CH$_3$CHO was also tentatively detected in NGC\,253 
    \citep{Meier15} and towards a galaxy at redshift z=0.89 \citep{Muller11}. Although not
much is known about this molecule, its origin seems to be linked to 
grain mantle destruction by shocks, and is found in massive
star-forming regions (e.g. \citealt{Chengalur03,Bennett05}). 

As commented in Sect.~\ref{madcuba}, we detected vibrationally
excited lines of HC$_3$N only in Arp\,220, which are listed in
Table~\ref{GaussParArp220}. They trace the hottest and densest molecular gas in the
inner regions of hot cores in star-forming regions \citep{DeVicente00,Martin-Pintado05}.

\subsection{Column densities and fractional abundances}
\label{columndensities}
 
 For those species not detected in a galaxy, we estimated
the 3$\sigma$ upper limit to the column density by assuming a
limit to the integrated area ($I$) given by

\begin{equation}
 I \le 3 \times rms \times  \sqrt{FWHM \times D_{\rm v}}
\end{equation}
 
where rms is the 1$\sigma$ noise level of the spectrum that contains the
line, $D_{\rm v}$ is the spectral resolution in velocity units indicated
in Sect.~\ref{sect.Obs} for each galaxy, and FWHM is
the average full width at half maximum of the Gaussian fits measured for
other species in the same galaxy.
As FWHM we generally used
95\,km\,s$^{-1}$ for M\,83, 200\,km\,s$^{-1}$ for NGC\,253,  100\,km\,s$^{-1}$
for  M\,82, 130\,km\,s$^{-1}$ for M\,51, 
240\,kms$^{-1}$ for NGC\,1068, 250\,km\,s$^{-1}$ for NGC\,7469,
390\,km\,s$^{-1}$ for Arp\,220, 
and 210\,km\,s$^{-1}$ for Mrk\,231. Note, however, that these values
may differ for some species.

We calculated fractional abundances of all the species (or their
upper limits in case they were not detected) relative to the
$^{12}$C$^{18}$O (hereafter C$^{18}$O)
column density,
which is significantly less affected by opacity effects than the main oxygen isotopologue
$^{12}$C$^{16}$O. We note, however, that C$^{18}$O seems to be
    enhanced in the two ULIRGs of our sample.  In Sect. ~\ref{ULIRGs}, we discuss how this
    might affect our comparison of abundances among galaxies.

\begin{table*}
\caption{Source averaged column densities [cm$^{-2}$]}
\centering
\begin{tabular}[!h]{lrrrrrrrr} 
\hline
\hline
Molecule &M\,83  &NGC\,253 &M\,82 & M\,51  \\

\hline
$^{13}$CO&  $(5.7\pm0.1)\times10^{16}$&$(1.88\pm0.03)\times10^{17}$& $(1.32\pm0.02)\times10^{17}$&$(2.1\pm0.1)\times10^{16}$\\
  C$^{18}$O &  $(1.10\pm0.03)\times10^{16}$&$(5.1\pm0.2)\times10^{16}$&  $(2.61\pm0.08)\times10^{16}$& $(5.2\pm0.4)\times10^{15}$ \\
   C$^{17}$O & $(1.9\pm0.2)\times10^{15}$&$(6.7\pm0.3)\times10^{15}$& $(1.6\pm0.3)\times10^{15}$& $(9.0\pm1.9)\times10^{14}$ \\
  HCN &  $(8.6\pm0.1)\times10^{13}$ &$(3.85\pm0.05)\times10^{14}$& $(2.90\pm0.06)\times10^{14}$& $(4.5\pm0.1)\times10^{13}$\\
 H$^{13}$CN & $\bf{(2.3\pm0.5)\times10^{12}}$&$(2.8\pm0.1)\times10^{13}$& $(5.0\pm0.6)\times10^{12}$& $(1.8\pm0.4)\times10^{12}$ \\
HNC       & $(3.13\pm0.09)\times10^{13}$&$(1.81\pm0.02)\times10^{14}$&$(1.16\pm0.03)\times10^{14}$& $(1.35\pm0.06)\times10^{13}$\\
HN$^{13}$C   & $\bf{(1.2\pm0.6)\times10^{12}}$ &$(1.5\pm0.1)\times10^{13}$&$\le 2.1\times10^{12}$& $\le 1.4\times10^{12}$\\
  HCO$^+$& $(4.6\pm0.1)\times10^{13}$&$(1.94\pm0.03)\times10^{14}$& $(2.55\pm0.06)\times10^{14}$& $(1.31\pm0.04)\times10^{13}$ \\
 H$^{13}$CO$^+$  &$\bf{(1.1\pm0.4)\times10^{12}}$&$(1.3\pm0.2)\times10^{13}$&$(6.9\pm0.3)\times10^{12}$& $\le6.6\times10^{11}$    \\
HC$^{18}$O$^+$ &-&$(4.3\pm0.2)\times10^{12}$&$(4.7\pm2.5)\times10^{12}$&-\\
HOC$^+$     & $\bf{(1.3\pm0.2)\times10^{12}}$&$(6.4\pm0.4)\times10^{12}$&$(3.6\pm0.2)\times10^{12}$& $\le 5.8\times10^{11}$  \\
  HCO        & $\le5.0\times10^{12}$&$(5.9\pm0.2)\times10^{13}$& $(5.1\pm0.6)\times10^{13}$&$\le 7.0\times10^{12}$\\
  C$_2$H  & $\bf{(5.0\pm0.1)\times10^{14}}$&$(2.82\pm0.04)\times10^{15}$& $(3.69\pm0.05)\times10^{15}$& $(1.6\pm0.1)\times10^{14}$ \\
  CN   & $\bf{(3.24\pm0.04)\times10^{14}}$&$(1.74\pm0.01)\times10^{15}$&$(1.02\pm0.02)\times10^{15}$& $(1.31\pm0.02)\times10^{14}$  \\
 $^{13}$CN     & $\le7.0\times10^{12}$&$(6.2\pm0.5)\times10^{13}$& $\le1.2 \times10^{13}$& $\le1.5\times10^{13}$\\ 
CH$_3$OH  &  $\bf{(1.9\pm0.2)\times10^{14}}$ &$(1.3\pm0.1)\times10^{15}$& $(2.07\pm0.08)\times10^{14}$& $(3.9\pm1.5)\times10^{13}$\\
CS      & $(5.7\pm0.1)\times10^{13}$&$(4.07\pm0.07)\times10^{14}$&$(2.50\pm0.05)\times10^{14}$& $(1.8\pm0.1)\times10^{13}$\\
 C$^{34}$S & $(7.7\pm1.0)\times10^{12}$& $(5.4\pm0.2)\times10^{13}$& $(1.50\pm0.08)\times10^{13}$& $(3.1\pm0.8)\times10^{12}$ \\
 $^{13}$CS   & $\le2.9\times10^{12}$&$(1.4\pm0.1)\times10^{13}$& $\le4.6\times10^{12}$ & $\le3.2\times10^{12}$ \\ 
HC$_3$N      &$\bf{(1.9\pm1.2)\times10^{13}}$&$(1.04\pm0.05)\times10^{14}$& $(4.6\pm0.4)\times10^{13}$& $\le4.4\times10^{12}$\\
  SO  & $\bf{(2.0\pm0.3)\times10^{13}}$ &$(1.7\pm0.3)\times10^{14}$& $(7.7\pm0.8)\times10^{13}$& $(1.1\pm0.3)\times10^{13}$ \\
SO$_2$    &$\le1.6\times10^{13}$ &$(3.6\pm0.6)\times10^{13}$& $\le2.0\times10^{13}$& $\le 1.3\times10^{13}$ \\  
NS      & $\le1.2\times10^{13}$ &$(3.8\pm1.9)\times10^{13}$&$(3.0\pm0.6)\times10^{13}$& $\le6.4\times10^{12}$  \\
 HNCO   & $(4.7\pm2.5)\times10^{13}$ &$(2.2\pm0.7)\times10^{14}$&  $\bf{(2.7\pm5.0)\times10^{13}}$& $\le 6.7\times10^{12}$  \\
 N$_2$H$^+$ & $\bf{(6.5\pm0.3)\times10^{12}}$ &$(3.97\pm0.07)\times10^{13}$& $(1.31\pm0.03)\times10^{13}$& $(4.1\pm0.3)\times10^{12}$  \\
  SiO  & $\bf{(2.4\pm0.6)\times10^{12}}$ &$(3.1\pm0.3)\times10^{13}$& $(4.4\pm0.6) \times10^{12}$& $\le 1.2\times10^{12}$ \\
CH$_3$CN  &$\bf{(3.4\pm0.9)\times10^{12}}$ &$(3.3\pm2.6)\times10^{13}$&  $(4.8\pm15.8)\times10^{12}$& $\le 1.8\times10^{12}$ \\
CH$_3$CCH      &$\bf{(9.4\pm5.7)\times10^{13}}$ &$(5.4\pm1.0)\times10^{14}$&  $(1.15\pm0.04)\times10^{15}$& $\le 5.3\times10^{13}$\\
c-C$_3$H$_2$    & $\le3.7\times10^{15}$ &$(1.05\pm0.06)\times10^{14}$& $(1.31\pm0.06)\times10^{14}$& $\le 1.7\times10^{15}$ \\
NH$_2$CN   &$\le3.2\times10^{12}$ &$(1.3\pm0.3)\times10^{13}$& $\le 3.9\times10^{12}$& $\le 4.0\times10^{12}$ \\
HOCO$^+$  &$\le6.4\times10^{12}$ &$(1.2\pm0.7)\times10^{13}$& $\le6.1\times10^{12}$& $\le 5.0\times10^{12}$\\
OCS &$\le 7.3\times10^{13}$&$(4.9\pm0.3)\times10^{14}$ & $\le 8.4\times10^{13}$& $\le 7.4\times10^{13}$\\
C$_2$S    & $\le7.6\times10^{12}$ &$(2.5\pm0.8)\times10^{13}$& $\le1.2\times10^{13}$& $\le 1.9\times10^{13}$ \\
H$_2$CS   &  $\le1.5\times10^{13}$ &$(3.8\pm0.9)\times10^{13}$& $\le 1.4\times10^{13}$& $\le 1.1\times10^{13}$ \\
HC$_5$N    &  $\le3.5\times10^{14}$&$\bf{(7.1\pm13.4)\times10^{13}}$& $\le3.8\times10^{14}$& $\le4.0\times10^{14}$ \\
CH$_3$CHO &$\le1.1\times10^{13}$&$\bf{(7.4\pm64.7)\times10^{13}}$& ${\bf(3.4\pm0.4)\times10^{13}}$& $\le1.5\times10^{13}$ \\
 \hline
\end{tabular}

\begin{list}{}{}

\item Molecules in boldface are new detections for a given 
  galaxy. Column densities have been calculated for molecules
      clearly detected, as well as for those tentatively detected. Upper
      limits are shown for species not detected.

\end{list}{}{}

\label{Nmol1}
\end{table*}

\begin{table*}
\caption{Source averaged column densities [cm$^{-2}$]}
\centering
\begin{tabular}[!h]{lrrrrrrrr} 
\hline
\hline
Molecule & NGC\,1068 & NGC7469 &Arp\,220& Mrk\,231  \\

\hline
$^{13}$CO& $(4.6\pm0.1)\times10^{17}$   & $(3.6\pm0.2)\times10^{16}$& $(6.6\pm0.4)\times10^{17}$ & $(5.6\pm0.9)\times10^{16}$\\
  C$^{18}$O & $(1.29\pm0.04)\times10^{17}$    &$(6.4\pm1.3)\times10^{15}$& $(6.2\pm0.4)\times10^{17}$&$(4.9\pm0.9)\times10^{16}$ \\
   C$^{17}$O & $\le 1.0\times10^{16}$   &$\le 3.9\times10^{15}$&$\le2.9\times10^{16}$&$\le2.2\times10^{16}$\\
  HCN &  $(2.45\pm0.04)\times10^{15}$    & $(1.04\pm0.08)\times10^{14}$&$(4.6\pm0.1)\times10^{15}$ &$(8.5\pm0.7)\times10^{14}$\\
 H$^{13}$CN &$(8.1\pm0.8)\times10^{13}$    &$\le 1.4\times10^{13}$&$\bf{(7.1\pm0.5)\times10^{14}}$&$(2.0\pm0.5)\times10^{14}$ \\
HNC       &$(7.0\pm0.1)\times10^{14}$    & $(4.2\pm0.5)\times10^{13}$ &$(3.2\pm0.2)\times10^{15}$ & $(2.5\pm0.7)\times10^{14}$\\
HN$^{13}$C   & $(2.4\pm0.6)\times10^{13}$     & $\le 1.6\times10^{13}$&$\le1.3\times10^{14}$&   $\le1.8\times10^{14}$ \\
  HCO$^+$& $(9.1\pm0.2)\times10^{14}$     &$(7.4\pm0.5)\times10^{13}$&$(1.12\pm0.06)\times10^{15}$&$(3.3\pm0.3)\times10^{14}$ \\
 H$^{13}$CO$^+$  & $\le 9.0\times10^{12}$& $\le 8.0\times10^{12}$& $(9.8\pm3.2)\times10^{13}$&   $\le9.1\times10^{13}$   \\
HC$^{18}$O$^+$&-&-&-&-\\
HOC$^+$     & $(1.6\pm0.4)\times10^{13}$    &  $\le 6.8\times10^{12}$& $\le5.6\times10^{13}$ & $\le6.2\times10^{13}$ \\
  HCO        & $(2.6\pm0.8)\times10^{14}$   & $\le 7.1\times10^{13}$& $\le5.7\times10^{14}$ & $\le8.0\times10^{14}$\\
  C$_2$H  & $(1.10\pm0.02)\times10^{16}$     & $(1.0\pm0.1)\times10^{15}$&$\bf{(2.2\pm0.1)\times10^{16}}$&$(3.4\pm0.7)\times10^{15}$ \\
  CN   & $(6.85\pm0.06)\times10^{15}$     & $(4.6\pm0.2)\times10^{14}$& $(6.1\pm0.2)\times10^{15}$&$(1.8\pm0.1)\times10^{15}$\\
 $^{13}$CN     & $\le 1.5\times10^{14}$    & $\le 6.6\times10^{13}$ & $\le5.8\times10^{14}$&  $\le1.4\times10^{14}$\\ 
CH$_3$OH  & $(2.4\pm0.2)\times10^{15}$  & $\le 1.8\times10^{14}$&$(5.6\pm0.6)\times10^{15}$&$\le2.6\times10^{15}$\\
CS      & $(7.6\pm0.2)\times10^{14}$     & $(8.1\pm1.0)\times10^{13}$ & $(3.36\pm0.08)\times10^{15}$ & $(3.5\pm0.8)\times10^{14}$\\
 C$^{34}$S & $(1.3\pm0.2)\times10^{14}$  & $\le 3.0\times10^{13}$&$(9.6\pm1.3)\times10^{14}$&$\le2.3\times10^{14}$ \\
 $^{13}$CS   & $\le 3.6\times10^{13}$    & $\le 3.0\times10^{13}$& $\le2.0\times10^{14}$ &  $\le2.0\times10^{14}$\\ 
HC$_3$N      & $(1.50\pm0.07)\times10^{14}$   &  $\le5.0\times10^{13}$& $(2.7\pm0.2)\times10^{15}$&  $(2.1\pm0.5)\times10^{14}$\\
  SO  & $(2.2\pm0.4)\times10^{14}$     &$\le 6.9\times10^{13}$& $\le4.4\times10^{14}$&$\le5.0\times10^{14}$ \\
SO$_2$    & $\le 7.4\times10^{15}$     &  $\le 1.1\times10^{14}$ & $\le 9.2\times10^{14}$& $\le1.1\times10^{15}$ \\  
NS      & $\le 9.6\times10^{13}$    & $\le 8.0\times10^{13}$&$\le 3.1\times10^{14}$ & $\le5.5\times10^{14}$  \\
 HNCO   &  $(3.2\pm0.4)\times10^{14}$     &$\le 6.0\times10^{13}$& $(1.8\pm0.9)\times10^{15}$ & $\le3.0\times10^{14}$  \\
 N$_2$H$^+$ & $(1.2\pm0.5)\times10^{14}$ & $\le 8.2\times10^{12}$& $(5.9\pm0.3)\times10^{14}$&$\le5.4\times10^{13}$ \\
  SiO  & $(6.8\pm0.8)\times10^{13}$     &$\le 1.5\times10^{13}$&$(6.3\pm0.7)\times10^{14}$& $\le1.7\times10^{14}$ \\
CH$_3$CN  & $(1.0\pm2.9)\times10^{14}$      &  $\le 1.7\times10^{13}$&$(9.2\pm1.7)\times10^{14}$&$\le9.1\times10^{13}$ \\
CH$_3$CCH      & $\le 5.7\times10^{14}$       & $\le 3.7\times10^{14}$  & $(3.7\pm0.8)\times10^{15}$&$\le4.0\times10^{15}$ \\
c-C$_3$H$_2$    & $\le2.7 \times10^{16}$     & $\le 3.9\times10^{15}$& $\le 4.4\times10^{16}$& $\le5.3\times10^{14}$  \\
NH$_2$CN   & $\le2.6\times10^{13}$    & $\le2.2\times10^{13}$ &  $\le 1.5\times10^{14}$&  $\le1.8\times10^{14}$ \\
HOCO$^+$ & $\le 7.7\times10^{13}$    &$ \le 4.4\times10^{13}$ & $\le 3.7\times10^{14}$ &   $\le3.5\times10^{14}$\\
OCS & $\le 8.9\times10^{14}$& $\le8.2\times10^{14}$&$\le 6.1\times10^{15}$&$\le9.2\times10^{15}$\\
C$_2$S    & $\le 1.2\times10^{14}$   & $\le 9.3\times10^{13}$& $\le 6.5\times10^{14}$&  $\le7.1\times10^{14}$ \\
H$_2$CS   & $\le1.6\times10^{14}$    &$\le 9.0\times10^{13}$ &$\le 6.6\times10^{14}$ &  $\le7.6\times10^{14}$ \\
HC$_5$N    & $\le5.4\times10^{15}$  & $\le 4.9\times10^{15}$& $\le4.0\times10^{16}$&  $\le3.9\times10^{16}$ \\
CH$_3$CHO & $\le4.6\times10^{14}$& $\le 1.2\times10^{14}$&$\le7.2\times10^{14}$&$\le1.3\times10^{15}$ \\
 \hline
\end{tabular}

\begin{list}{}{}

\item Molecules in boldface are new detections in a given galaxy.  Column densities have been calculated for molecules
      clearly detected, as well as for those tentatively detected. Upper
      limits are shown for species not detected.

\end{list}{}{}

\label{Nmol2}
\end{table*}

\section{Key molecular species}
\label{keyspecies}

\begin{figure*}
\centering
\includegraphics[width=1.3\textwidth, angle=90 ]{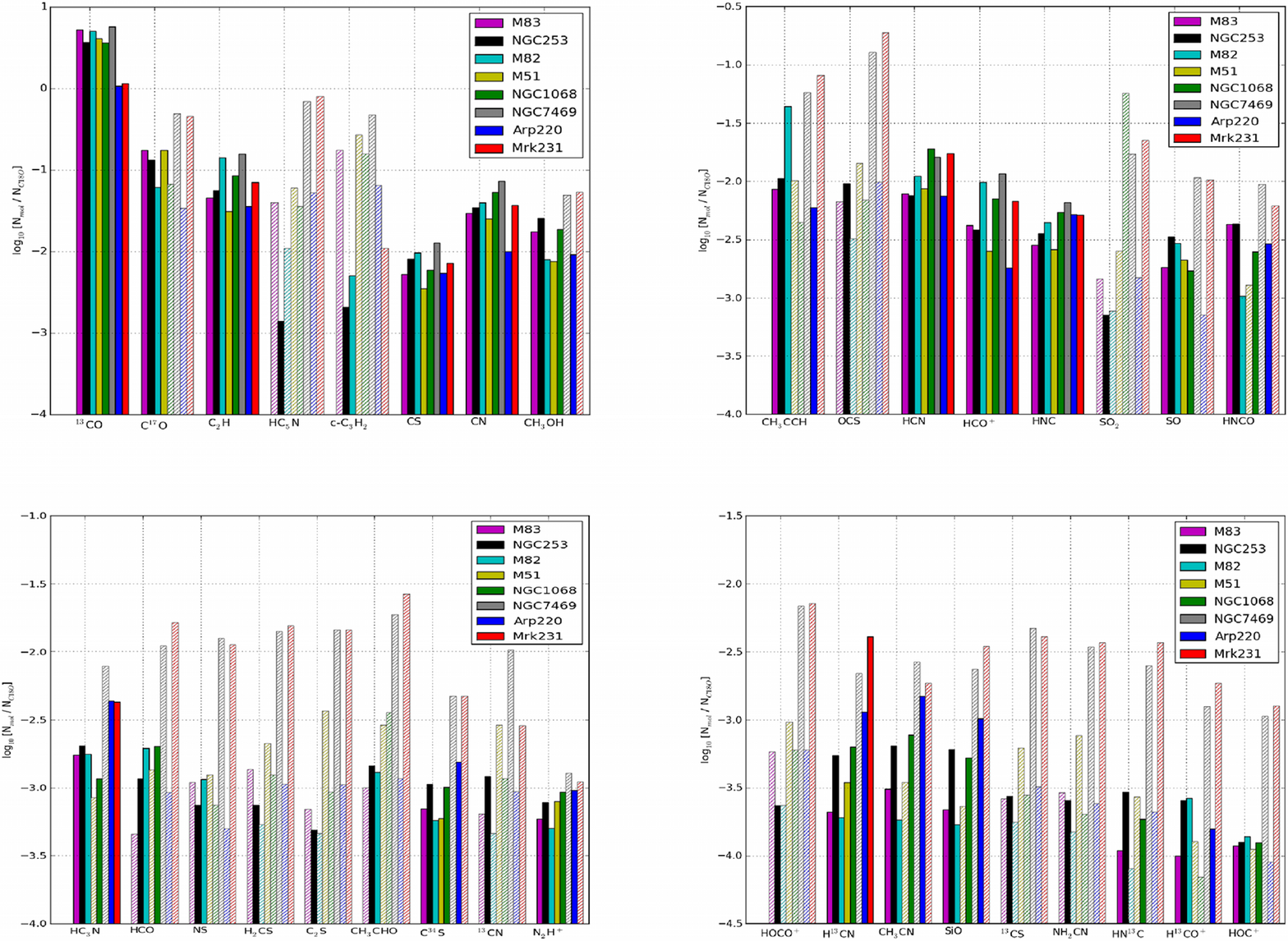}
\caption{Abundances of thirty-four species relative to C$^{18}$O for the
  eight galaxies of our sample. Dashed histograms indicate upper limits.}
\label{abundances}
\end{figure*}

Figure~\ref{abundances} shows the comparison of thirty-five molecular
abundances among the eight galaxies in the sample.
Table~\ref{ratios} lists the most relevant column density ratios used
for the discussion in this section and the following. Based on these results, we indicate here the species 
presenting the highest abundance contrasts among our galaxy
sample. These molecules are potentially good candidates for further
high-resolution follow up observations. We note that we only discuss here species where the abundance
difference is at least a factor of three.

\begin{itemize}

\item CH$_3$CCH is detected in M\,83, NGC\,253, M\,82, and Arp\,220, but not in the
AGNs of our sample. M\,82 has the maximum CH$_3$CCH abundance and, in fact, no other galaxy has
  been found to have such a high value (at least relative to H$_2$, C$^{34}$S
  and C$^{18}$O, \citealt{Aladro11a,Aladro11b,Aladro13}). The other
  galaxies where this species was detected 
  have similar values, five to seven times lower than that of M\,82. Interestingly,
  CH$_3$CCH was not detected in NGC\,1068, NGC\,7469, M\,51, and Mrk\,231, which
  supports the idea that this molecule is highly contrasted between
  starbursts and AGNs \citep{Aladro13}. The upper limits to the
  abundance of M\,51 and NGC\,1068 are significant, and show that
  the difference in abundance between M\,82 and NGC\,1068 is about one
  order of magnitude. 

In the Galaxy, CH$_3$CCH  was found to be very abundant in
regions
of prolific star formation, where its column
    density 
    is substantially higher
    ($N_{\rm CH_3CCH}\sim10^{14-15}$\,cm$^{-2}$, \citealt{Churchwell83,Miettinen06}), than in dark clouds,
    ($N_{\rm CH_3CCH}\sim10^{13}$\,cm$^{-2}$, \citealt{Irvine81,Markwick05}). Our values
    for M\,82, M\,83, NGC\,253 and Arp\,220 are of the same
    order as the ones cited above for high-mass star-forming
    cores. However, given the different scales
        in  extragalactic and galactic sources, the comparison of the
        column densities is not straightforward. In
    addition, it was recently found that CH$_3$CCH is enhanced in PDRs in
    the Horsehead nebula
    as compared to dense cores \citep{Guzman14}. High
    temperatures could help to understand high abundances of CH$_3$CCH in
  these objects, as it may form after the desorption of its
  chemical precursors from grain mantles. In fact, rotational
  temperatures derived for Orion-KL, Sgr\,B2, DR\,21 and  many other
  similar sources are in the range [20-50]\,K \citep{Churchwell83,Miettinen06}.

As previously claimed by \citet{Aladro13}, and in agreement with  our
    results here, the abundance  of CH$_3$CCH seems
to be enhanced in 
starburst galaxies. It is a known tracer of relatively dense gas (the critical densities for the two transitions
observed here are $10^4-10^5$\,cm$^{-3}$). Its emission is
more widespread than that of other similar molecules, such as CH$_3$CN, and can also
be
found in regions outside the cores of the molecular clouds
\citep{Aladro11a}. The response of CH$_3$CCH
to the presence of UV fields and shocks has not been studied in detail
in extragalactic sources. Consequently, the chemistry of CH$_3$CCH is still uncertain, but it might well be that
the paths for its production 
are more effective in the presence of UV radiation fields created by
massive star formation (see \citealt{Aladro11b} and references
therein). Supporting this idea, recent interferometric maps of
    several species in the central 
    half kiloparsec of NGC\,253 show that the emission of CH$_3$CCH has a similar
    morphology to other PDR tracers, such as C$_2$H \citep{Meier15}. Thus, it seems
    that, apart from being a dense gas tracer,
    CH$_3$CCH can also be considered a PDR tracer.

\item HC$_3$N and H$^{13}$CN  show clearly enhanced abundances in Arp\,220 and
  Mrk\,231 with respect to the rest of galaxies in our sample. This could be related to
  larger amounts of dense gas and warm dust present in ULIRGs (as compared to less
  obscured galaxies, \citealt{Lindberg11}), where the dust would prevent their
  photo-dissociation. On average, the
  differences between the ULIRGs and the other galaxies is a factor of three for
  HC$_3$N and seven for H$^{13}$CN, although the maximum differences
  between individual galaxies are of factors four and
  twenty-one for HC$_3$N and H$^{13}$CN respectively.

\item CH$_3$CN and SiO have maximum abundances in Arp\,220, which are a factor of 
  $\sim$7 higher than those of M\,82, which
  shows the minimum in both cases. Both species are found to be tracers of dense
  gas and shocks, although enhanced SiO abundances may also be related
  to X-ray chemistry in AGNs \citep{Burillo10,Aladro13}. The species are
  photo-dissociated in the external layers of molecular clouds if
  strong UV fields are present. This scenario is in agreement with
  both molecules having high abundances in Arp\,220, which probably has more
  dense gas and dust, and a minimum in M\,82, as it is heavily pervaded by UV fields.  

\item C$_2$H. Previous observational studies found high abundances of
  this species in 
  PDRs of starburst galaxies (e.g. \citealt{Meier05,
    Fuente05,Aladro11b}). Together with M\,82, the
  highest abundances of C$_2$H in our sample are also 
  those of NGC\,7469 and NGC\,1068, while M\,51 presents the lowest
  value, followed by Arp\,220 and M\,83. There is a factor five of
  difference between the abundances of M\,82 and M\,51. 
 In the case of NGC\,1068 and NGC\,7469, significant emission is likely to be arising from the circumnuclear star
forming rings,  as in NGC~1097 \citep{Martin14b}.

\item CN and HCN abundances stand out in NGC\,7469, NGC\,1068, Mrk\,231, and
  M\,82, being the difference between these galaxies and the rest
  of the sample larger in HCN. In agreement with previous observations (e.g. \citealt{Aalto02,Aalto08}) and model predictions \citep{Lepp96, Sternberg95,Meijerink05}, our data indicate that CN and HCN could be effectively produced in the X-ray and UV-field dominated regions expected in AGN and starburst galaxies.

\item HNCO is tentatively detected for the first
time in M\,82. This species is
claimed to be a good indicator of the evolution of nuclear
starbursts in galaxies, with abundances decreasing toward the last
evolutionary 
stages of the starbursts (in those late phases HNCO should be
efficiently destroyed by means of reactions involving the abundant UV
photons and/or ions).
HNCO remained undetected in previous studies of M\,82 \citep{Martin09,Aladro11b}. Our new tentative HNCO line detections in M\,82 have
intensities of $\sim$3\,mK, which are higher than the upper limit of $<$1.3\,mK previously estimated by \citet{Martin09}. Our HNCO
abundances in M\,82 are four times lower than in
NGC\,253 and M\,83.
This contrast is smaller than
  that found by
  \citet{Martin09} for HNCO/C$^{34}$S, which shows differences of nearly two orders of
  magnitude, but goes in the same direction. Therefore, our
    observations agree with HNCO being a good tracer of the
    evolutionary stage of starburst galaxies.

\item C$^{32}$S/C$^{34}$S. 
Both $^{32}$S and $^{34}$S are believed to be a product of
hydrostatic or explosive oxygen-burning in massive stars and Type Ia
supernovae (see \citealt{Chin96} and references therein). However,
not much is known apart from
some Galactic observations which indicate that $^{32}$S/$^{34}$S 
increases with  the galactocentric radius
\citep{Chin96}. In external
galaxies, more data is needed to
know how this 
ratio varies from source to source.

We found that the C$^{32}$S\,$(2-1)$/C$^{34}$S\,$(2-1)$ line ratio is five times higher in
M\,82 (with the maximum value of  of 21.5 in our sample) than in
  Arp\,220 (minimum value of 4.5, similar to that derived with SO by
  \citealt{Martin11}). 
This points to a higher CS opacity in Arp\,220 where there is much more molecular gas.
As a reference value,
  C$^{32}$S/C$^{34}$S in the Galactic centre (GC),
  the local ISM, and the Solar system is
  $\sim$22, \citep{Frerking80}. M\,82 has a similar value 
  to the GC, which is $\sim$3 times higher than that of
  NGC\,253 (7.6). More sophisticated estimates by \citet{Martin10}
  and \citet{Henkel14}  give 
  values close to the local interstellar value  for NGC\,253 ($>$16).

 NGC\,1068,  M\,51, NGC\,253, and M\,83 have C$^{32}$S/C$^{34}$S
 $J=2\rightarrow$1 line ratios of $6-8$, well
 below the GC and M\,82, and closer to
  the value in Arp\,220.  Though these ratios point out towards the
  degree of processing of the ISM in the galactic nuclei, the values
should be taken with care, as opacity on the main isotopologue might
play an important role in these regions  (\citealt{Martin10,Henkel14} and
references therein). Opacity becomes less of an issue for LIRGs, based on
the $^{13}$CO/C$^{18}$O sample measured by \citet{Costagliola11}.

\end{itemize}

\section{Chemical caracterisation of our sample}
\label{discussion}

\begin{table*}
\caption{Relevant column density ratios used for the discussion. }
\centering
\begin{tabular}[!h]{lcccccccc} 
\hline
\hline
Molecular ratio & M\,83  &NGC\,253 &M\,82 &M\,51&  NGC\,1068 & NGC7469  &Arp\,220& Mrk\,231 \\

\hline
 HNCO / C$^{18}$O & $4.3\times 10^{-3}$  &$4.3\times 10^{-3}$ &$1.0\times 10^{-3}$ & $\le 1.3\times 10^{-3}$  & $2.5\times 10^{-3}$ &  $\le9.4\times 10^{-3}$ & $2.9\times 10^{-3}$  &$\le6.1\times 10^{-3}$ \\
  CH$_3$OH / C$^{18}$O  & $1.7\times 10^{-2}$  &$2.5\times 10^{-2}$ &$7.9\times 10^{-3}$ & $ 7.5\times 10^{-3}$  & $1.9\times 10^{-2}$ &  $\le2.8 \times 10^{-2}$ & $9.0\times 10^{-3}$  &$\le5.3\times 10^{-2}$ \\
  SiO / C$^{18}$O &  $2.2\times 10^{-4}$  &$6.1\times 10^{-4}$ &$1.7\times 10^{-4}$ & $\le5.0 \times 10^{-5}$  & $5.3\times 10^{-3}$ &  $\le 2.3\times 10^{-3}$ & $1.1\times 10^{-3}$  &$\le3.5\times 10^{-3}$ \\
  CH$_3$CN / C$^{18}$O  & $3.1\times 10^{-4}$  &$6.5\times 10^{-4}$ &$1.8\times 10^{-4}$ &$\le1.8\times 10^{-3}$  & $7.8\times 10^{-4}$ &  $\le2.6\times 10^{-3}$ & $1.5\times 10^{-3}$  &$\le1.8\times 10^{-3}$ \\
 CH$_3$CCH / C$^{18}$O  &  $8.5\times10^{-3}$  & $1.0\times10^{-2}$   &  $4.4\times10^{-2}$  &
 $\le1.0\times10^{-2}$  & $\le4.4\times10^{-3}$  & $\le5.8\times10^{-2}$   &$6.0\times10^{-3}$    & $\le8.2\times10^{-2}$  \\
 SO / C$^{18}$O   &  $1.8\times 10^{-3}$  &$3.3\times 10^{-3}$ &$3.0\times 10^{-3}$ & $ 2.1\times 10^{-3}$  & $1.7\times 10^{-3}$ &  $\le1.1\times 10^{-2}$ & $\le7.1\times 10^{-4}$  &$\le1.0\times 10^{-2}$ \\
  c-C$_3$H$_2$ / C$^{18}$O  & $\le3.4\times 10^{-1}$  &$2.0\times 10^{-3}$ &$5.1\times 10^{-3}$ & $\le3.3\times 10^{-1}$  & $\le2.1\times 10^{-1}$ &  $\le6.1\times 10^{-1}$ & $\le7.1\times 10^{-2}$  &$\le1.1\times 10^{-2}$ \\
  HCO / C$^{18}$O & $\le4.5\times 10^{-4}$  &$1.2\times 10^{-3}$ &$2.0\times 10^{-3}$ & $\le1.3\times 10^{-3}$  & $2.0\times 10^{-3}$ &  $\le1.1\times 10^{-2}$ & $\le9.2\times 10^{-4}$  &$\le1.6\times 10^{-2}$ \\
  C$_2$H / C$^{18}$O   &  $4.5\times 10^{-2}$  &$5.5\times 10^{-2}$ &$1.4\times 10^{-1}$ & $ 3.1\times 10^{-2}$  & $8.5\times 10^{-2}$ &  $ 1.6\times 10^{-1}$ & $3.5\times 10^{-2}$  &$6.9\times 10^{-2}$ \\
  N$_2$H$^+$ / C$^{18}$O &$5.9\times 10^{-4}$  &$7.8\times 10^{-4}$ &$5.1\times 10^{-4}$ & $ 7.9\times 10^{-4}$  & $2.5\times 10^{-3}$ &  $\le1.3\times 10^{-3}$ & $9.5\times 10^{-4}$  &$\le1.1\times 10^{-3}$ \\
  HNC / C$^{18}$O  &  $2.8\times 10^{-3}$  &$3.5\times 10^{-3}$ &$4.4\times 10^{-3}$ & $ 2.6\times 10^{-3}$  & $5.4\times 10^{-3}$ &  $ 6.6\times 10^{-3}$ & $5.2\times 10^{-3}$  &$5.1\times 10^{-3}$ \\
HCN / C$^{18}$O  &$7.8\times 10^{-3}$  &$7.5\times 10^{-3}$ &$1.1\times 10^{-2}$ & $ 8.6\times 10^{-3}$  & $1.9\times 10^{-2}$ &  $ 1.6\times 10^{-2}$ & $7.4\times 10^{-3}$  &$1.7\times 10^{-2}$ \\
CN / C$^{18}$O  &$2.9\times 10^{-2}$  &$3.4\times 10^{-2}$ &$3.9\times 10^{-2}$ & $ 2.5\times 10^{-2}$  & $5.3\times 10^{-2}$ &  $ 7.2\times 10^{-2}$ & $9.8\times 10^{-3}$  &$3.7\times 10^{-2}$ \\
 HCN / HCO$^+$  &  1.9    &   2.0   &    1.1  &    3.4  &     2.7     &  1.4    &   4.1   &    2.6 \\
  CN / HCN   & 3.8 &     4.5   &   3.5  &    2.9   &   2.8   &  4.4   &   1.3  &    2.1  \\
HCN /  CS & 1.5 &    0.9   &   1.2  &    2.5  &    3.2  &    1.3    &  1.4   &   2.4  \\
\hline
\end{tabular}

\label{ratios}
\begin{list}{}{}
\item The errors of the ratios range in general between 2\% and 63\%  and have
  an average value of 14\%. Exceptions to these values are the ratios involving HNCO
      and CH$_3$CN in M\,82, as their column densities have large errors associated.
\end{list}{}{}
\end{table*}

\begin{figure*}[h!]
       \centering
     \includegraphics[angle=90,width=0.8\textwidth]{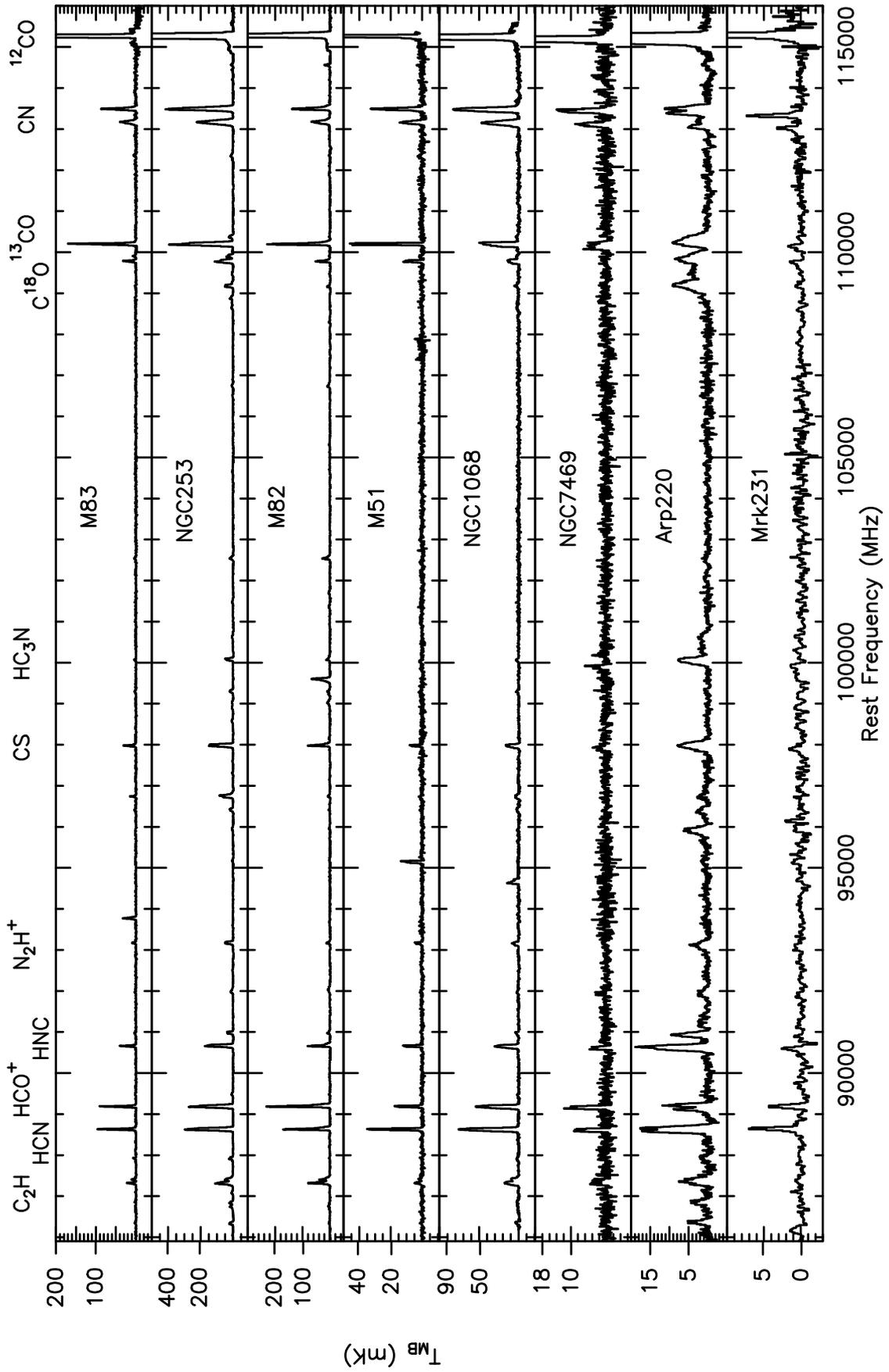}
      \caption{Spectra of all eight galaxies between 86 and 116\,GHz. The
        strongest molecules are indicated. The faint detected species
        are shown in
        Figs.~\ref{abundances-faint-1} and ~\ref{abundances-faint-2}. The spectra refer to the 
            rest frame adopting the velocities in Table~\ref{galaxies}.
}
\label{AllSurveys1}
\end{figure*}

\begin{figure*}
\centering
\includegraphics[width=1.3\textwidth, angle=90 ]{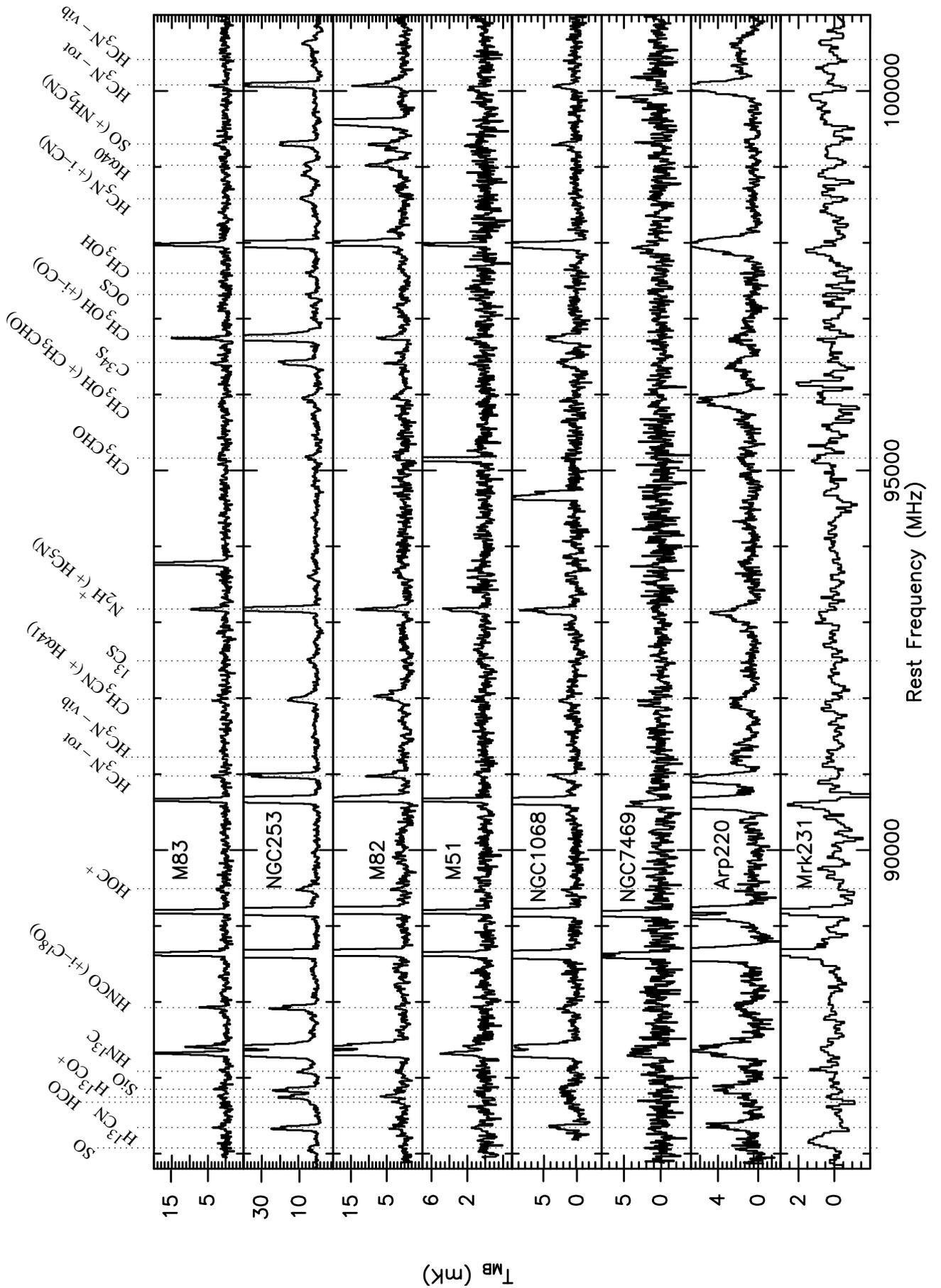}
\caption{Zoomed spectra between 86\,GHz and 100\,GHz where the faint lines detected are
  highlighted. Lines
in brackets were not detected in all the galaxies (typically they were only
detected in NGC\,253). For details about the detections see Appendix~\ref{LongTables}. The spectra refer to the 
            rest frame adopting the velocities in Table~\ref{galaxies}.}
\label{abundances-faint-1}
\end{figure*}
\begin{figure*}
\centering
\includegraphics[width=1.3\textwidth, angle=90 ]{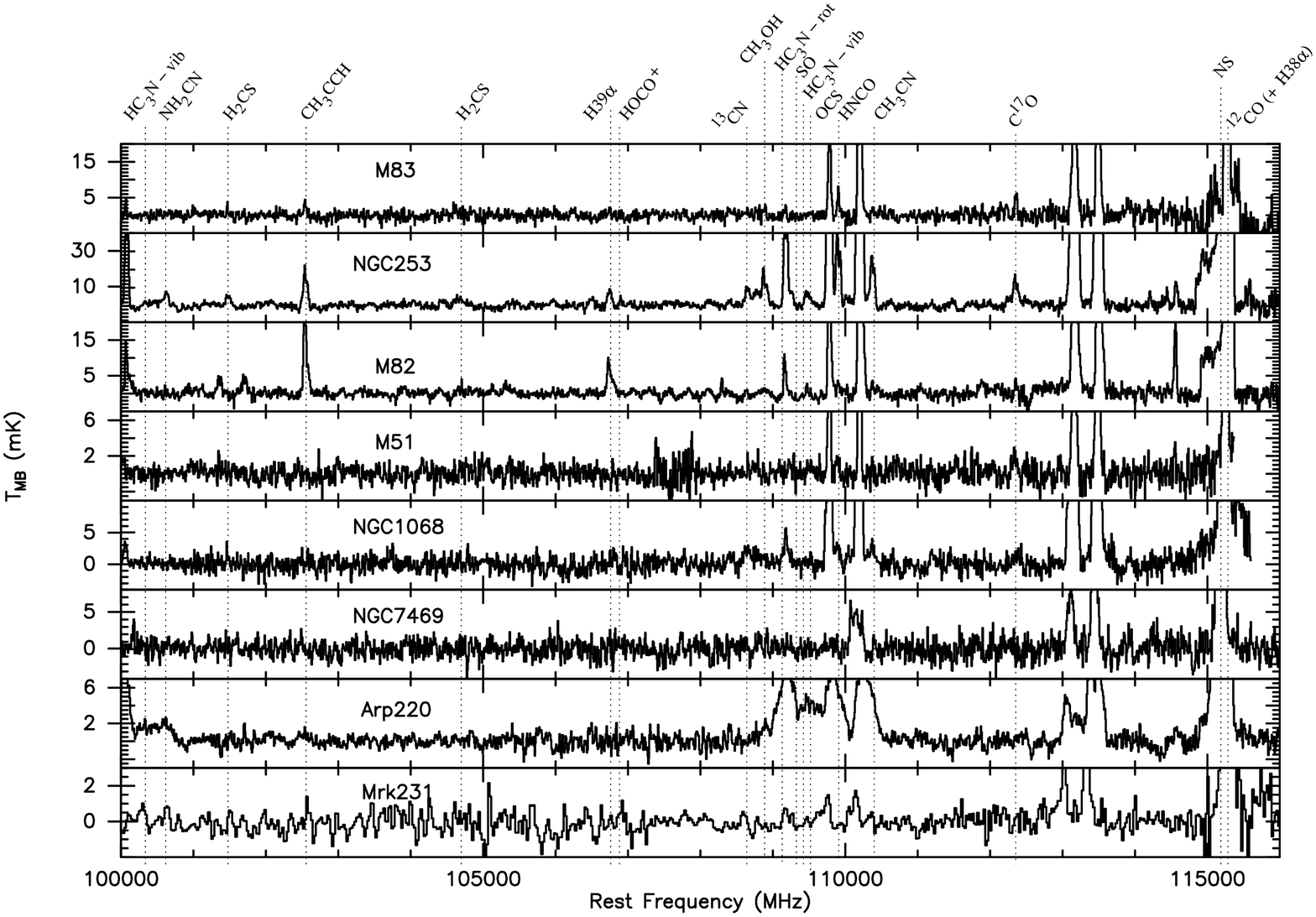}
\caption{Continued zoomed spectra between 100\,GHz and 116\,GHz where the faint lines detected are
  highlighted. Lines
in brackets were not detected in all the galaxies (typically they were only
detected in NGC\,253). For details about the detections see
Appendix~\ref{LongTables}.  The spectra refer to the 
            rest frame adopting the velocities in Table~\ref{galaxies}.}
\label{abundances-faint-2}
\end{figure*}

\subsection{Starburst galaxies}
\label{SB}

The molecular gas in the nuclear region of M\,83 is rich in some
shock tracers and species formed in dust grains, such
as HNCO and CH$_3$OH (Fig.~\ref{abundances} and Table~\ref{ratios}), that are easily
released through grain mantle evaporation by shock waves. Although the
abundances of these two species are high relative to C$^{18}$O 
($10^{-3}$ and $10^{-2}$ for HNCO and CH$_3$OH respectively), other shock and
dense gas tracers are just marginally detected (e.g. SiO, CH$_3$CN,
and SO). No H$\alpha$ recombination lines were detected in M\,83,
unlike in the more active starbursts (see SFR in
    Table~\ref{galaxies}) M\,82 and NGC\,253, in spite of all three
    galaxies having similar beam filling factors. This suggests that
    big and medium-size HII regions and OB star clusters are
    relatively scarce in M\,83.
Accordingly, the PDR-tracers c-C$_3$H$_2$ and
HCO were not detected.

As for  NGC\,253, groups of OB stars create photo-dissociated regions
in the outskirts of the molecular clouds. This is confirmed by
previous detections of PDR tracers, such as HOC$^+$, CO$^+$, HCO, and c-C$_3$H$_2$
\citep{Martin09}, which  are easily formed by ion-molecule reactions
in gas phase (e.g. \citealt{Hollis83,Thaddeus85}). 
However, the gas in NGC\,253 is mainly characterised by the presence
of many shock
tracers and species formed in the icy mantles of dust grains. HNCO, CH$_3$OH, and SO are a factor of two more
abundant than in M\,83. 
Also, some other faint shock tracers have been only
detected in this galaxy, such as HOCO$^+$ and NS, as well as
 the sulphur bearing species
SO$_2$. H$_2$CS, and C$_2$S (\citealt{Martin03,Martin05},) which got confirmed here (see Fig.~\ref{abundances}). 

The gas in the NE lobe of M\,82 is strongly
affected by UV fields which shape its chemical composition (e.g. \citealt{Aladro11b}). Molecular clouds are highly ionized and
fragmented \citep{Fuente08} and, according to our results, the
PDR tracers c-C$_3$H$_2$, HCO, and HOC$^+$ have slightly higher abundances than in
NGC\,253 (but just a factor of two or less. See Table~\ref{ratios}). The species that shows the largest difference between M\,82 and
the other two starburst galaxies is CH$_3$CCH, being in the former five
times more abundant (see Sect.~\ref{keyspecies} for more details).
Still, some shock tracers, such as
CH$_3$OH, SiO, and HNCO, are detectable in the cores of the molecular clouds that
are well shielded from the UV radiation \citep{Martin06a}. However, their abundances are
four times lower than in NGC\,253, indicating that, although present,
shocks do not represent the main heating mechanism of the molecular
gas as compared to UV radiation.

\subsubsection{A possible chemical scenario of starburst evolution}
As commented in Sect.~\ref{galaxysample}, M\,83, NGC\,253, and M\,82
can be taken as examples of starbursts in a  early,
intermediate, and late stages of evolution. If so, it
may be possible to link their evolutionary stages to the chemical
composition of their molecular gas. This evolutionary scenario
    has been progressively fine-tuned by several authors \citep{Burillo02, Martin06b, Aladro11b}. 
An early starburst is expected to be chemically characterised by the
absence of most PDR tracers (HCO, c-C$_3$H$_2$)  and H$\alpha$
recombination lines, as well as the presence of only
the most common
shock-tracers (HNCO, CH$_3$OH, SiO). Physically, first
shocks could be caused by close encounters
or mergers between galaxies, or collisions of giant molecular
cloud complexes (for example, following the bar potential toward the centre of the
galaxy, \citealt{Binney91}). 

In an intermediate starburst stage, with similar characteristics to
NGC\,253, the stellar population would contain more recently formed
young massive stars and supernovae events that 
ionise the outer layers of the molecular clouds, as seen by the detection
of the PDR tracers c-C$_3$H$_2$, HCO and HOC$^+$. During this phase,
 PDR tracer abundance enhancements are still smaller than those caused
 by shocks, which would reach their maximum influence over the ISM at
 this point. Molecules such as CH$_3$OH,
HNCO, and SO have higher abundances than in a previous (and later)
phase. Other shock tracers would be also detectable,
specifically  HOCO$^+$ and some sulphur
bearing species, such as H$_2$CS, OCS, SO, SO$_2$ and C$_2$S
(see \citealt{Martin03,Martin05} and Fig.~\ref{abundances}), whose
abundances are enhanced during  the last evolutionary stages of hot cores.

Finally, an advanced starburst such as that found in M\,82, would be
characterised by super-winds created by massive stars and supernovae,
as well as a larger population of high-mass X-ray binaries, which
provide a modest X-ray emission, needed to understand the abundance of
some molecular species, such as CO$^+$ \citep{Spaans07,Chiang11}. Stellar winds enrich
the ISM with both PDR tracers (HCO, HOC$^+$, c-C$_3$H$_2$, and CH$_3$CCH) and shock-tracers
(CH$_3$OH, SiO, OCS, SO), the first ones being more abundant, since UV
fields are more intense than in previous stages. Less abundant, or
more complex, shock tracers and sulphur species would be easily photo-dissociated by the
radiation fields, and thus would become less abundant than in previous
starburst phases or not detectable at all.

\subsection{Seyfert galaxies}
\label{AGNs}
Among the galaxies in our sample, M\,51, NGC\,1068, and
NGC\,7469 host an AGN (the first two are classified as Sy\,2,
and the latter as Sy\,1, \citealt{Osterbrock93}).
A detailed study devoted to our results in NGC\,1068  has
been presented by \citet{Aladro13}. The CNDs of these galaxies are surrounded by arms of active
star forming regions at $\sim1-2$\,kpc distance from their
centres. Given the angular resolution of the IRAM 30\,m telescope
at 3\,mm wavelengths ($22''-29''$), our data embrace the CNDs as well as the
surrounding regions, so some contribution from the star forming
rings is expected in our data. Other observations of NGC\,1068 
indicate that some molecules show significant emission from the starburst ring ($<25\%$ for
SiO\,$(2-1)$ and $<30\%$ for HCO$^+(1-0)$, \citealt{Usero04}), while high
  transitions of the dense gas tracers HCN, HCO$^+$ and CS are hardly
  detected outside the CND \citep{Burillo14}. Recently, Takajima et al (in prep.) 
      carried out a similar molecular line survey of NGC\,1068 with
      the Nobeyama 45\,m telescope. Thanks to the smaller beam size of that
      telescope at $\lambda$=3\,mm, their observations can separate
      the emission of the CND and that of the starburst ring, and show
      that the emission of some species such as C$_2$H and N$_2$H$^+$
      come mainly from the latter.
Unfortunately, there are
  no similar estimations for NGC\,7469
  and M\,51 calculated for the IRAM 30\,m beam at the 3\,mm
  wavelength of our survey.

CH$_3$CCH and c-C$_3$H$_2$ were not detected in any of the AGNs, while HCO and
HOC$^+$ are only tentatively seen in NGC\,1068, and have similar
abundances than in the
starburst galaxies. Regarding shocks and dust grain chemistry, most of the
tracers (such as OCS, HNCO, or SO) are only detected in
NGC\,1068. Shocks seem to be affecting part of the gas in the CND of this galaxy 
\citep{Krips11,Viti14}.
 CH$_3$OH is also
detected in M\,51 (not in NGC\,7469), but with a clearly lower
abundance than in NGC\,1068.

Among others, we examined the following relative abundances: HCN/C$^{18}$O, CN/C$^{18}$O,
HCN/HCO$^+$, CN/HCN, and HCN/CS (Table~\ref{ratios}). The last three
were  proposed
as discriminators between AGNs and starburst activities based on
    two-dimensional plots of line ratios
\citep{Kohno01,Meijerink07,Krips08,Izumi13}. 
Using our data, HCN/CS, HCO$^+$/HCN,
and CN/HCN, either by themselves and plotting one ratio
    versus other, do not show a correlation with the
galactic nuclear activity of our sample. The most probable reason for the
     lack of trend between all these line ratios and AGNs (versus
     starburst galaxies) is likely the low angular resolution of
     the IRAM 30\,m telescope, which is not sufficient to separate the CND
     and the surrounding starburst regions. However, we do see that the HCN  abundance is clearly enhanced in
AGNs, and the reason for that could be 
a high temperature-driven chemistry \citep{Izumi13}, or shocked
    regions at a few hundred parsecs from the super massive black
    holes due to outflowing material \citep{Martin14b}.
Interestingly,  HCN/C$^{18}$O, CN/C$^{18}$O, C$_2$H/C$^{18}$O, and HNC/C$^{18}$O in M\,51
 are closer to the starburst results than to the other
 AGNs (Fig.~\ref{abundances}). 


SiO  is tentatively detected in NGC\,1068 but not
in M\,51 and NGC\,7469.  In the case of NGC\,7469, where lines
    are weaker, we cannot
    conclude from the upper limit to the SiO abundance whether the gas
  chemistry is different.
This species usually traces shocks in 
Galactic and extragalactic sources
\citep{Martin-Pintado92,Martin-Pintado97, Burillo00,Burillo01,Riquelme10,
Takano14}. However, some studies
(based on observational data or on chemical models)
point to SiO as a tracer of X-rays and cosmic rays in the
centres of the Milky Way and NGC\,1068
\citep{Martin-Pintado00, Usero04, Aladro13}. If that is the case in all AGNs, its detection
in the other two Seyferts of our sample would be somehow expected (if
SiO was strong enough to be detectable). A study devoted to this molecule
is necessary to better understand how it is formed under the different
regimes (i.e. shock-dominated versus X-ray dominated gas).

Using the IRAM
30\,m telescope, \citet{Watanabe14} carried out a molecular line survey,
in the 3\,mm and 2\,mm bands, outside the
nucleus of M\,51. They observed two regions
in a spiral arm of the galaxy.
Their position P1 has twice the SFR of
position P2. Both SFRs are quite low, in the range [0.02-0.05]$M_\odot$yr$^{-1}$.
This allows us, for the first time, to compare the nuclear
    molecular gas composition with that outside the central region. Comparing the fractional
    abundances (with respect to C$^{18}$O) of the twenty-two species in
    common between our work and Watanabe's, we find that the chemical
    composition of the gas is fairly the same, except for a few
    cases.  H$^{13}$CO$^+$, HNCO, and c-C$_3$H$_2$ are detected in the P1 position,
    but not in the centre. On the other hand, we detect H$^{13}$CN, C$^{34}$S,
    and SO, which are not seen in the P2 position. Apart from that,
  the rest of the molecules have similar abundances, within a factor
  of $<$2 in the three positions. One would expect that the chemistry of the gas might change  between the nucleus, affected by higher star formation and the
  proximity of the black hole, and the spiral arms. However, given the
  beam size of the  IRAM 30\,m at these frequencies and the source
  sizes of the three positions, the emission might be so diluted that
  the differences are cancelled out. Further investigations with
  interferometers are needed to better understand the change of the
  gas chemistry as a function of galactocentric radius. 


\subsection{ULIRGs}
\label{ULIRGs}

Independently of their powering source,
the ULIRGs of our sample show
some chemical differences with respect to the AGNs and starburst galaxies
described in the previous sections, as well as some
dissimilarities between Arp\,220 and Mrk\,231 themselves.

The HNC\,$(1-0)$ line intensity in Arp\,220  slightly exceeds that of
HCN\,$(1-0)$ by a factor of 1.3, while for Mrk\,231  we obtain HNC/HCN
$(1-0)$\,=\,0.6.  This particular high ratio in Arp\,220 was already
noticed by \citet{Huettemeister95} and \citet{,Aalto07}, and seems to
be due to an overabundance of HNC in this galaxy,  a consequence of 
either mid-IR pumping of the lines, or a chemistry  driven by X-ray
 dominated regions \citep{Meijerink07,Aalto07}. A priori, we cannot rule out any
 of the two scenarios with our data, since we would need several HCN and HNC
 transitions in order to estimate the H$_2$ column and volume densities,
 indispensable to address the situation. However, we dare to favour 
 IR pumping effects, given that the chemical composition of
 Arp\,220 is not similar to the AGNs of our sample (XDR scenario), as we
 describe below.

 CH$_3$CN and 
SiO  are
relatively abundant in Arp\,220 (few$\times 10^{-3}$ with respect to
C$^{18}$O), but are not detected in Mrk\,231. 
On the other hand, HCO$^+$, H$^{13}$CN, and CN are four times more abundant in  Mrk\,231 than in
Arp\,220. 
Our results suggest that the Arp\,220 chemistry
resembles more the starburst galaxies of our sample dominated by shocks,  with
high abundances of HC$_3$N, CH$_3$OH, HNCO, CH$_3$CN and SiO relative to C$^{18}$O (10$^{-2,-3}$),  but no
PDR tracer detections (c-C$_3$H$_2$, HCO, HOC$^+)$, except for
CH$_3$CCH (just tentatively detected). This would support the
infrared pumping of HNC in the galaxy at the expense of a XDR environment, as
mentioned before.
Mrk\,231, in contrast, is characterised by
high abundances of CN, HCN and H$^{13}$CN relative to C$^{18}$O (10$^{-2,-3}$), more similar to those found in
NGC\,1068 and NGC\,7469.

Vibrationally excited lines from HC$_3$N, arising from dense ($n\ge
10^6$\,cm$^{-3}$) and hot ($T_{\rm ex}= 190\pm20$\,K) cores, are detected in
Arp\,220.  They are a factor 4-5 fainter than the corresponding
rotational transitions. Assuming that this ratio may be similar 
 for the rest of the galaxies, the expected vibrational intensities in Mrk\,231 would be $\sim$0.2\,mK, which is
well below our rms level. Thus, we cannot discard lower level
vibrational emission in Mrk\,231. However, vibrationally excited HC$_3$N
lines would be observable in NGC\,253 and M\,82, but are not
detected. This clearly points to a different excitation between normal
starbursts and ULIRGs.

There are also common features between the
Mrk\,231 and Arp\,220 spectra as
compared to the rest of the sample. In agreement with previous studies
\citep{Greve09,Martin11,Henkel14} $^{13}$CO and C$^{18}$O are equally
abundant in both ULIRGs. This may be due to
high opacities  of the $^{13}$CO lines, to an intrinsic underabundance of $^{13}$CO 
in luminous infrared galaxies \citep{Taniguchi99}, or/and to a
C$^{18}$O overabundance due to the strong starbursts
\citep{Matsushita09, Martin11}. High opacities of  $^{13}$CO in these
two galaxies seem unlikely, given the weakness of the $^{13}$CO$(1-0)$
lines relative to$^{12}$CO$(1-0)$ (they are $\ge$20 times
fainter). An
overabundance of $^{18}$O in Arp\,220 has been recently found
by \citet{Gonzalez14}. However, the $^{12}$C/$^{13}$C and $^{16}$O/$^{18}$O line
    ratios obtained for Arp\,220 and Mrk\,231 by \citet{Henkel14} do
    not favour this last scenario.
It can be seen in Fig.~\ref{abundances} that the $^{13}$CO/C$^{18}$O ratios in Mrk\,231 and
Arp\,220 are a factor of $\sim$4 lower than in the rest of the
galaxies. We obtain a $^{12}$CO/$^{13}$CO\,$(1-0)$ relative
    intensity of 21$\pm$7, which is a factor two lower than that
    obtained by \citet{Greve09} for Arp\,200 (43$\pm$10). We note that
  our $^{12}$CO\,$(1-0)$ and $^{13}$CO\,$(1-0)$ lines were observed simultaneously with the
  EMIR receiver of the 30m telescope, while \citet{Greve09} used
  older data to derive this ratio, which may be less accurate.

In addition,  HC$_3$N/C$^{18}$O ratios are clearly
higher  in the ULIRGs than in the more nearby starbursts and AGNs observed
by us. If  C$^{18}$O is overabundant in ULIRGs, then HC$_3$N should be it as well. This effect was already observed in other LIRGs/ULIRGs, and the reason could be a
higher amount of dense gas and warm dust  that protects the species from
being photo-dissociated \citep{Aalto07,Costagliola11,Lindberg11}.
Thus, HC$_3$N seems to be a well
suited species to study the activity in ULIRGs,
regardless of the powering source in their nuclei. Also common to Arp\,220 and Mrk\,231 is
the outstanding H$^{13}$CN/C$^{18}$O ratio, as
previously commented in Sect.~\ref{keyspecies}.  Again, if
    C$^{18}$O is overabundant in Arp\,220 and Mrk\,231,  it
  only makes more
    clear the difference in abundances of HC$_3$N and H$^{13}$CN between the
    ULIRGs and the rest of the sample. On the other hand, if  C$^{18}$O has
    similar values in all the galaxies, it results in an even  bigger contrast.

It is also interesting to compare ULIRGs, characterised by strong
interactions between galaxies, with isolated galaxies, where the
chemical and physical properties are the result of intrinsic and secular
mechanisms. In CIG\,638, a good representative of an isolated
galaxy (see \citealt{Verley07} for the criteria used to define these objects), the C$_2$H/HCN and C$_2$H/HNC integrated intensity ratios were found to be
interestingly high (1.0 and 4.3 respectively), either because the galaxy has
a lesser amount of dense gas, or/and due to an overproduction of C$_2$H \citep{Martin14}. The C$_2$H/HCN
ratios for Arp\,220 and Mrk\,231 are a factor of 3 and 4 (respectively) lower than in
CIG\,638. Moreover, the C$_2$H/HNC  integrated intensity ratio  is
even more contrasted, being 9 and 5 times (respectively) lower in the ULIRGs than
in the isolated galaxy. This points to a very likely difference
     in the gas composition between
these two types of galaxies, with the ULIRGs containing more dense gas
(associated with their nuclear starbursts), than the isolated
galaxies. On the other hand, the rest of the galaxies of our sample have a C$_2$H/HNC ratio in the
range $\sim[0.7-1.5]$ which, according to the results derived
by \citet{Martin14}, agree with the expected values for
``normal'' (local) starbursts and AGNs. These results are on average well below the
CIG\,638 and above the Arp\,220 and Mrk\,231 values.

\section{Summary and conclusions}
\label{conclusions}

Using the IRAM 30\,m telescope, we performed a molecular line survey
towards the circumnuclear regions of eight active
galaxies, covering the frequency range $\sim$[86-116]\,GHz. The sample is
composed by three starburst galaxies, M\,83, M\,82, and NGC\,253,
three AGN galaxies
with starburst influence, NGC\,1068, M\,51, and NGC\,7469, and two
ULIRGs, Mrk\,231 and Arp\,220. The column densities
  of  twenty-seven species (or their upper limits when they are not detected) were
calculated assuming LTE and a single source size for each galaxy based
on previous available interferometric maps. The studied species include ten $^{13}$C, $^{18}$O, $^{17}$O, and $^{34}$S bearing
isotopologues. We compared the abundances, with respect to C$^{18}$O,
among the eight galaxies to characterise the average molecular
    gas composition of each kind of galaxies.
Our results are
summarised in Figs. ~\ref{AllSurveys1} and~\ref{abundances} and Tables~\ref{Nmol1},~\ref{Nmol2}, and ~\ref{ratios}.

The comparison among the galaxies allowed us to find
out the key species that show the most contrasted abundances, relative to C$^{18}$O, among the
sources. CH$_3$CCH is only detected in starbursts and is a factor of  one order of magnitude more abundant in M\,82 (the galaxy with the
highest abundance) than in NGC\,1068 (not detected). This molecule
seems to be associated with massive star formation, and might be a good
tracer of PDRs. H$^{13}$CN, CH$_3$CN and SiO have a
peak of abundance in Arp\,220, which is twenty-one (H$^{13}$CN), and
seven (CH$_3$CN, SiO) times
respectively higher than in M\,82 (minimum abundance). We find high
C$_2$H abundances in M\,82, NGC\,1068 and NGC\,7469, being in NGC\,7469 five times higher than in
M\,51 (minimum). A large part  of the C$_2$H emission in the AGNs is
    expected to come from the starburst rings that surround the
    CNDs. HNCO shows
abundance variations among the starburst galaxies of a factor of four. This
latter result supports previous claims of HNCO as a good tracer of the
starburst evolutionary state \citep{Martin09}. Other species, such as
HCN and HC$_3$N, show smaller, though still significant, abundance differences among galaxies by factors of
three to four. 
All the cited molecules constitute good 
candidates to study the chemistry in galaxy nuclei. 

Regarding the starburst  galaxies, the molecular gas of M\,83 is characterised by relatively high
abundances (with respect to C$^{18}$O, 10$^{-2}$-10$^{-4}$) of some shock tracers,
such as CH$_3$OH, HNCO and SiO, as well as the absence of both H$\alpha$ recombination
lines and the PDR tracers c-C$_3$H$_2$ and HCO. In NGC\,253 shocks, created by
 molecular cloud collisions and stellar winds, 
dominate the heating of the gas, as shown by the presence of several other shock- and
dust grain-tracers (e.g. SO, SO$_2$, OCS, HOCO$^+$). In addition, the molecular cloud complexes are
ionised by OB clusters, which translates into high
intensities of H$\alpha$ recombination
lines, as well as moderate abundances of c-C$_3$H$_2$, HOC$^+$, and HCO (10$^{-3}$-10$^{-4}$).
Lastly, M\,82 represents a galaxy  where UV fields are very intense and create
large PDRs. H$\alpha$ recombination lines, as well as 
c-C$_3$H$_2$, HCO, HOC$^+$  and CH$_3$CCH are brighter than in the other starbursts
of the sample. CH$_3$OH and HNCO are still
observable in the cores of the molecular clouds, although their abundances are
clearly lower than in NGC\,253 and M\,83, due to
the lack of shielding from UV photons. We explore a possible scenario where M\,83, NGC\,253
and M\,82 are considered templates of galaxies in young, intermediate and late
phases of starbursts, and follow the evolution of the 
gas composition along time.

As for the AGNs (NGC\,1068, NGC\,7469, M\,51 and Mrk\,231), 
HCN/HCO$^+$ and HCN/CS
are quite low in NGC\,7469, similar to the values in starburst galaxies; HCN/C$^{18}$O  and
CN/C$^{18}$O, C$_2$H/C$^{18}$O, and HNC/C$^{18}$O are also much the same in
starbursts and in the low luminosity AGN of M\,51.  
The lack of molecular trends in our AGN galaxies, as
compared to the starbursts, seems to 
be due to their CNDs not being spatially resolved by the IRAM 30\,m
beam. Remarkably, CH$_3$OH, SiO and HNCO are only detected in
NGC\,1068.

Finally, the ULIRGs Mrk\,231 and Arp\,220 show common spectral
features, such as the almost identical $^{13}$CO and
C$^{18}$O column densities, and higher H$^{13}$CN/C$^{18}$O
and HC$_3$N/C$^{18}$O abundance ratios  than the rest of the galaxies. Nevertheless, both ULIRGs
differ in some line detections that seem to point out to different 
mechanisms at work in their nuclear regions. In particular,  Arp\,220
shows emission of HC$_3$N vibrationally excited lines that arise from
warm gas ($T_{\rm vib}=190\pm20$\,K), and shock tracers such as
CH$_3$OH, HNCO, and SiO. This, together with the
 lack of PDR tracers such as HOC$^+$, HCO, and c-C$_3$H$_2$, makes
 Arp\,220 resemble
 the spectrum of a giant hot core-like starburst. On
the other hand,  Mrk\,231 has higher abundances of HCN, H$^{13}$CN, 
and HCO$^+$, like the values of NGC\,1068 and NGC\,7469, thus
being more similar to an AGN-dominated galaxy.

\begin{acknowledgements}
This work has been partially funded by MICINN grants
AYA2010-21697-C05-01 and FIS2012-39162-C06-01, and
Astro-Madrid (CAM S2009/ESP-1496).
We wish to thank Jacques Le Bourlot and the Paris' Master of Astrophysics students for sharing their M\,82
data with us, observed with the IRAM 30\,m telescope at 3\,mm
wavelengths. These data were added to our own observations.
\end{acknowledgements}

\onecolumn
\clearpage

\begin{appendix}
\section{Simulated fit results for individual spectra.}
\label{LongTables}

\begin{table*}[t!]
\caption{Results for {\bf{M\,83}}}

\centering
\begin{tabular}[!h]{lrrrrrrrrrrrrr} 
\hline
\hline
Line	&	Frequency	&	$I$ &    $V_{\rm LSR}$   	  &    $\Delta V$ 		& $T^{\rm peak}_{\rm MB}$ 		& Comments	\\
	&	MHz		&	K\,km\,s$^{-1}$		&  km\,s$^{-1}$	  &   km\,s$^{-1}$	& mK	&\\	
\hline
H$^{13}$CN\,$(1-0)$ &	86340.2	&$ 0.25\pm 0.04$ &  $503.4\pm6.8$	& $64.7\pm16.0$&	3.55& hf, t	\\
H$^{13}$CO$^+$\,$(1-0)$&	86754.3&$0.21\pm  0.06$&   510& 95	&2.07	& t\\
SiO\,$(2-1)$	&	86847.0&$0.24\pm 0.06$	&   510&  95	&2.40& t \\
HN$^{13}$C\,$(1-0)$ &	87090.8	& $0.11\pm  0.05$	& 510	&  95	& 1.11		& t \\ 
C$_2$H\,$(1-0)$	&	87316.9&$4.01\pm 0.08$	&  $504.3\pm1.3$	& $91.7\pm3.2$&  22.62 	& hf \\ 
HNCO\,$(4_{0,4}-3_{0,3})$	&	87925.2&$0.69\pm 0.09$&  510	& 95&6.79& m	\\
HCN\,$(1-0)$	&	88631.8&$10.18\pm 0.13$&  $501.7\pm0.8$& $95.5\pm1.8$&97.47& hf	\\ 
HCO$^+$\,$(1-0)$ &89188.6&$9.30\pm  0.15$&  $502.8\pm1.0$& $95.0\pm2.4$		&   90.96	&	\\ 
HOC$^+\,(1-0)$&89487.4&$0.29\pm0.04$  	&  $516.3\pm7.7$& 95	&  2.87		&t	\\
HNC\,$(1-0)$	&90663.6&$4.12\pm0.08$&  $505.3\pm1.3$ 	& $93.0\pm3.0$& 41.45		&	\\
HC$_3$N\,$(10-9)$	&  	90979.0&$0.44\pm 0.05$&  $512.8\pm6.4$& 95	& 4.34		&	\\
CH$_3$CN$\,(5_k-4_k)$&91987.0&$0.24\pm0.06$	& 510& 95&2.33& m, t	\\ 
N$_2$H$^+$\,$(1-0)$ &	93173.7&$1.17\pm0.05$& 510	& 95& 13.32& m  \\
C$^{34}$S\,$(2-1)$ &	96412.9&$0.42\pm0.05$& 510 	& 95	& 	4.19	&\\
CH$_3$OH\,$(2_k-1_k)$&96741.4&$1.08\pm0.05$& $504.1\pm1.9$	& $57.5\pm4.5$&17.13	& m	\\
CS\,$(2-1)$	&	97981.0&$3.25\pm0.07$	&$501.9\pm1.4$&95&32.04	& 	\\
SO\,$(2_1-3_2)$&	99299.9&$0.45\pm0.05$&  $503.3\pm9.3$	& $120.0\pm22.1$	& 	3.49	&t	\\
HC$_3$N\,$(11-10)$&  100076.4	&$0.36\pm0.04$	&  $512.8\pm6.4$& 95& 	3.51	& t\\ 
CH$_3$CCH\,$(6_k-5_k)$ &102548.0 & $0.44\pm0.05$& $521.5\pm22.2$&95 &4.23 &m, t\\
CH$_3$OH\,$(0_0-1_{-1})$&108893.9&$0.20\pm0.01$& $504.1\pm1.9$	& $57.5\pm4.5$&3.28&  t	\\
HC$_3$N\,$(12-11)$&  	109173.6	&$0.26\pm0.03$& $512.8\pm6.4$& 95& 2.61		& t\\ 
C$^{18}$O\,$(1-0)$&109782.2&$3.40\pm0.07$ 	& $502.2\pm1.3$	& $94.7\pm3.1$&33.65&	\\ 
HNCO\,$(5_{0,5}-4_{0,4})$	&109905.8	&$0.70\pm0.09$&510	& 95&6.87& m\\ 
$^{13}$CO\,$(1-0)$&110201.4&$17.29\pm0.24$	& $502.2\pm0.9$& $97.8\pm2.1$ 	&165.00&  \\ 
CH$_3$CN\,$(6_k-5_k)$&110383.5	&$0.27\pm0.06$& 510	& 95&2.65&m, t\\ 
C$^{17}$O\,$(1-0)$&112359.3&$0.63\pm0.06$	& 510& 95&6.24& t \\ 
CN\,$(1_{0,1}-0_{0,1})$&113191.3& 	$6.31\pm0.06$& $500.7\pm0.9$	& $118.4\pm2.2$&31.14&m \\ 
CN\,$(1_{0,2}-0_{0,1})$&113491.0& $12.66\pm0.13$& $500.7\pm0.9$	&$118.4\pm2.2$ & 91.81	&m \\
$^{12}$CO\,$(1-0)$ &115271.2&$225.92\pm2.85$	& $504.4\pm1.3$& $79.6\pm2.8$& 1940.18	& 	\\
\hline
\end{tabular}

\begin{list}{}{}
\item[] Remarks: $(b)$ blended line; $(m)$
  multi-transition line; $(hf)$ hyperfine transition; $(t)$ tentative
  detection. Parameters without
      errors were fixed when fitting the lines.
\end{list}{}{}

\label{TableM83}
\end{table*}


\begin{center}
\begin{longtable}{lrrrrrrrrrrrrrrrr}
\caption[]{Results for {\bf{NGC\,253}}}
\label{TablaNGC253}\\
\hline
\hline
Line	&	Frequency	&	$I$ &    $V_{\rm LSR}$   	  &     $\Delta V$ 		& $T^{\rm peak}_{\rm MB}$ 		& Comments	\\
	&	MHz		&	K\,km\,s$^{-1}$		&  km\,s$^{-1}$	  &   km\,s$^{-1}$	& mK		&		\\
\hline

\endfirsthead

Continued from previous page\\
\hline
\endhead

Continue in next page
\endfoot

  See caption of Table~\ref{TableM83} for details.\\
\endlastfoot

NH$_2$CN\,$(4_{1,3}-3_{1,2})$&   80504.6    &$0.70\pm0.15$&	250&	180	 & 3.68 &t\\
C$_2$S\,$(6_7-5_6)$    &   81505.2    &$0.66\pm0.18$ &	250 &	200 &3.08&t\\
HC$_3$N \,$(9-8)$     &   81881.5 & $7.37\pm0.15$ &	250 & 180 & 38.41 &\\
c-C$_3$H$_2$\,$(2_{0,2}-1_{1,1})$&   82093.6 &	$1.66\pm0.07$&	250 & 180 &8.67 &\\
HC$_5$N\,(31-30) &   82539.3 &	$0.99\pm0.45$ &	240& 180 &5.15 &t\\
c-C$_3$H$_2$\,$(3_{1,2}-3_{0,3})$&   82966.2&	$0.76\pm0.03$&	250 & 180 &3.94 &t\\
CH$_3$OH\,$(5_{-1}-4_0)$ &84521.2&$2.52\pm0.12$&$249.4\pm5.1$&$217.7\pm11.9$&10.9\\
OCS\,$(7-6)$       &   85139.1 &$1.39\pm0.08$ & 	250&200 &6.53 &b\\
HC$^{18}$O$^+$\,$(1-2)$  &   85162.2 &	$0.84\pm0.04$&	250& 200 &3.93 &b\\
HC$_5$N\,(32-31) &   85201.6 &	$0.92\pm0.41$ &	240& 180& 4.78&b\\
c-C$_3$H$_2$\,$(2_{1,2}-1_{0,1})$&   85338.9 &$5.99\pm0.26$ & 250& 180 &31.09 &\\
CH$_3$CCH\,$(5_K-4_K)$       &   85457.3&$2.66\pm0.29$ & 250& 200 & 12.39 & m\\
HOCO$^+$\,$(4_{0,4}-3_{0,3})$    &   85531.5 &$0.38\pm0.22$ &	250& 180 &1.98 &t\\
H42$\alpha$    &   85688.4&	 - &	 230&  180 &8 &b\\
SO\,$(2_2-1_1)$             &   86094.0 &	$0.88\pm0.07$ &	250& 180 &4.60 &\\
H$^{13}$CN\,$(1-0)$ &	86340.2	& 	$4.71\pm 0.13$	&$250.1\pm3.2$	&  $181.5\pm7.4$	& 24.25	& hf 	\\
HCO\,$(1_{1,0}-0_{1,0})$&	86670.8		&$1.92\pm 1.17$	&250& 200	&4.68	& b, hf	& \\
H$^{13}$CO$^+$\,$(1-0)$&	86754.3&$3.81\pm0.44$&  250	& 200	&17.86		& b \\
SiO\,$(2-1)$	&	86847.0		& 	$4.76\pm0.44$	&  250	&  200	&	22.30	&  \\
HN$^{13}$C\,$(1-0)$ &	87090.8& $2.09\pm0.12$	&   250	& 200	& 9.81		&  \\ 
C$_2$H\,$(1-0)$	&	87316.9&$34.64\pm 0.45$	&  $246.9\pm1.9$& $186.8\pm3.8$&   106.10	& hf \\ 
HC$_5$N\,(33-32)&   87863.9 &	$0.84\pm0.38$ &	 240&  180 &4.39 &t, b\\
HNCO\,$(4_{0,4}-3_{0,3})$	&	87925.2&$5.79\pm1.73$&  250	& 200& 	26.99& m, b	\\
HCN\,$(1-0)$	&	88631.8		& 	$68.37\pm 0.72$	&  $249.1\pm1.3$	& $199.7\pm3.1$	&  305.20		& hf	\\ 
HCO$^+$\,$(1-0)$ &	89188.6		& 	$57.92\pm0.63$	&  $254.8\pm1.4$ 	& $199.7\pm3.4$	&   266.77	&	\\ 
HOC\,$^+(1-0)$	&	89487.4		&$2.13\pm0.11$	&  250	& 200	&  	10.02	&	\\
HNC\,$(1-0)$	&	90663.6		& 	$35.45\pm0.27$	&  $238.6\pm0.9$ 	& $185.1\pm2.2$	& 177.46		&	b\\
HC$_5$N\,(34-33)&   90526.2 &	$0.76\pm0.34$ &	 240& 180&3.99 &t, b\\
HC$_3$N\,$(10-9)$	&  	90979.0&$8.42\pm0.18$	&  250& 180	& 43.90		&	\\
CH$_3$CN$\,(5_k-4_k)$&91987.0&$3.59\pm2.66$	& 250	& 180&18.55	& m, b	\\ 
H41$\alpha$   &   92034.4&	- &	230& 180 &5 &b\\
$^{13}$CS\,$(2-1)$ 	&	92494.3	& $0.99\pm0.07$& 250	& 200	&  4.65		& 	\\ 
N$_2$H$^+$\,$(1-0)$ &	93173.7&$10.64\pm0.19$& $235.7\pm2.0$	&180&54.89  & m, b  \\
HC$_5$N\,(35-34) &   93188.5 &	$0.69\pm0.31$ &	 240& 180&3.59 &t, b\\
CH$_3$CHO\,$(5_{1,5}-4_{1,4})$\,A         & 93580.9 &	$0.52\pm4.05$ &	250& 200 &2.44 & t\\
CH$_3$CHO\,$(5_{1,5}-4_{1,4})$\,E         & 93595.2 &$0.52\pm4.03$ &	250& 200&2.43 &t \\
C$_2$S$\,(7_8-6_7)$ 	&	93870.1		& $0.68\pm0.19$&  250	& 200	&  	3.18	& t	\\ 
CH$_3$OH$\,(8_0-7_1)$&	95169.4& $0.98\pm0.12$& $233.3\pm10.6$& $171.8\pm24.2$	&5.38	& 	\\ 
HC$_5$N\,(36-35)&   95850.7 &	$0.61\pm0.28$ &	 240&  180&3.21 &t, b\\
CH$_3$OH$\,(2_1-1_1)$&	95914.3		&$0.81\pm0.01$	& 250	& 200	&	3.80	& b	\\ 
CH$_3$CHO\,$(5_{0,5}-4_{0,4})$\,E         & 95947.4 &	$0.65\pm5.03$&250&  200 &3.04 & b, t\\
CH$_3$CHO\,$(5_{0,5}-4_{0,4})$\,A         & 95963.4 &	$0.65\pm5.06$ &250&  200&3.06 & b, t\\
C$^{34}$S\,$(2-1)$ &	96412.9		& $	4.39\pm0.13$	& 250	& 200	& 	20.58	& 	\\
CH$_3$OH\,$(2_k-1_k)$&	96741.4&$17.14\pm0.26$ &	250	& 200	& 78.40& m, b	\\
OCS\,$(8-7)$&   97301.2 &$1.38\pm0.08$&	 250& 200&6.47& \\
CH$_3$OH\,$(2_1-1_1)$&97582.8&$0.83\pm0.01$	& 250& 200& 	3.89	& 	\\
CS\,$(2-1)$	&	97981.0		& 	$33.98\pm0.46$	&$234.5\pm1.7$ 	& 	$ 195.2\pm4.1 $	& 162	& 	\\ 
HC$_5$N\,(37-36) &   98512.9 &$0.54\pm0.24$ &	240&  180&2.83 &t\\
H40$\alpha$ &	99023.0&$1.45\pm0.50$& 237	& 180& 7		& 	\\
SO\,$(2_3-1_2)$&   99299.9 &	$4.20\pm0.32$ &250&180 &21.91&b\\
NH$_2$CN\,$(5_{1,5}-4_{1,4})$&  99311.2   &	$1.07\pm0.23$&	250&	180	& 5.58&m, b \\
NH$_2$CN\,$(5_{0,5}-4_{0,4})$&  99972.7   &	$1.06\pm0.23$&	250&	 180	& 5.51&\\
HC$_3$N\,$(11-10)$&  100076.4	&$9.02\pm0.19$&  250& 180&47.02& \\ 
NH$_2$CN\,$(5_{1,4}-4_{1,3}) $  &  100629.5   &$1.09\pm0.24$&	250	 &180&5.66 &m \\
HC$_5$N\,$(38-37)$ &  101175.1 &	$0.48\pm0.21$ &	240& 180&2.49 &t\\
H$_2$CS\,$(3_{1,3}-2_{1,2})$   &  101477.6   &	$0.58\pm0.12$	 &	 250	 &	180	&3.02 & t\\
CH$_3$CCH\,$(6_{k}-5_{k})$    &  102548.0   &$3.89\pm0.42$	 &	250	 &	 200	& 18.10 & m\\
H$_2$CS\,$(3_{0,3}-2_{0,2})$   &  103040.2   &	$0.83\pm0.18$	 &	250	 &	180	&4.33 & t\\
HC$_5$N\,$(39-38)$ &  103837.3 &$0.41\pm0.19$ &	240&  180&2.16 &t\\
SO$_2$\,$(3_{1,3}-2_{0,2})$&  104029.4   &$0.56\pm0.32$&	 240	 &	180	&2.90& t\\
H$_2$CS\,$(3_{1,2}-2_{1,1})$     &  104617.0  &	$0.60\pm0.13$	 &	 250	 &	180& 3.15& t\\
C$_2$S$\,(8_9-7_8)$ 	&106349.7& 	$0.61\pm0.17$&  250	& 200	&  	2.87	& t	\\ 
H39$\alpha$    &   106737.4&$1.75\pm0.34$ &	220& 181 &9 &\\
HOCO$^+$\,$(5_{0,5}-4_{0,4})$  &  106913.6  &$0.59\pm0.34$	 &	250	 &	180	& 3.07& t \\
$^{13}$CN\,$(1_{2,1}-0_{1,0})$ & 108651.3	&$1.08\pm0.09$&250	& 200	& 9.72	& m, b \\
$^{13}$CN\,$(1_{1,1}-0_{1,1})$ & 108657.6	&$1.08\pm0.09$&250	& 200& 9.72& m, b\\
$^{13}$CN\,$(1_{2,2}-0_{1,1})$ & 108780.2	&$1.94\pm0.16$&250	& 200& 8.92& m \\
CH$_3$OH\,$(0_{0}-1_{-1})$ & 108893.9	&	$3.50\pm0.05$	&250 & 200& 16.39		&  \\
HC$_3$N\,$(12-11)$  	&  109173.6	& 	$9.12\pm0.19$&  250& 180& 47.56		& \\ 
SO\,$(3_2-2_1)$&	109252.2	& $1.40\pm0.11$& 250& 180&7.28& t\\ 
OCS\,$(9-8)$  	&  109463.1	& $1.20\pm0.07$&  250& 200	& 	5.65	& b, t\\ 
C$^{18}$O\,$(1-0)$&	109782.2	& $22.49\pm0.57$	& $258.3\pm3.3$	&$204.5\pm7.8$	& 	102.65	&	\\ 
HNCO\,$(5_{0,5}-4_{0,4})$	&	109905.8	& $7.62\pm2.28$& 250& 200		& 	35.55	& m \\ 
$^{13}$CO\,$(1-0)$&	110201.4	& $81.16\pm0.96$	& $259.9\pm1.6$	& $200.8\pm3.6$ 	& 	370.67	&  \\ 
CH$_3$CN\,$(6_k-5_k)$&	110383.5	& $5.03\pm3.73$	& 250	& 180& 25.94		& m\\ 
C$^{17}$O\,$(1-0)$  	&  112359.3	& $3.15\pm0.14$	&  $269.9\pm5.7$	& 200	& 14.77	& \\ 
CN\,$(1_{0,1}-0_{0,1})$	&113191.3	& $47.93\pm0.32$	& $249.6\pm0.8$& $193.2\pm2.0$	 	& 184.30		&m \\ 
CN\,$(1_{0,2}-0_{0,1})$	&113491.0	& $95.53\pm0.62$	&$249.6\pm0.8$ & $193.2\pm2.0$	& 425.50	&m\\
NS\,$(3_{1,-1}-2_{1,1})$& 115153.9	&$1.80\pm0.90$	& 250& 180& 	8.86	& m, b, t \\
$^{12}$CO\,$(1-0)$ &	115271.2	& 	$1069.30\pm8.60$	&$260.9\pm0.1$& $210.6\pm0.1$		& 4769.30	& b	\\
H38$\alpha$    &   115274.4& -&	 230&  180 &7 &b\\
NS\,$(3_{1,1}-2_{1,-1})$	& 115556.2	& $1.80\pm0.90$	&250	& 180		& 8.32& m, t\\	
\hline



\end{longtable}
\end{center}

\begin{table*}[h!]
\caption{Results for  {\bf{M\,82}}}

\centering
\begin{tabular}[!h]{lrrrrrrrrrrrrrrr} 
\hline
\hline
Line	&	Frequency	&	$I$ &    $V_{\rm LSR}$   	  &      $\Delta V$ 		& $T^{\rm peak}_{\rm MB}$ 		&Comments	\\
	&	MHz		&	K\,km\,s$^{-1}$		&  km\,s$^{-1}$	  &   km\,s$^{-1}$	& mK &		&		\\	
\hline
c-C$_3$H$_2$\,$(3_{2,2}-3_{1,3})$ &84727.7&$0.143\pm0.06$&300&100&1.34& t\\
HC$^{18}$O$^+$\,$(1-0)$ &85162.2&$0.41\pm0.06$&300&100&3.82&b\\
c-C$_3$H$_2$\,$(2_{1,2}-1_{1,0})$ &85338.9& $3.36\pm0.13$&300&100&31.16& \\
CH$_3$CCH\,$(5_{\rm k}-4_{\rm k})$ & 85457.3&$2.59\pm0.08$&$314.6\pm1.9$&100&23.94 & m\\
H42$\alpha$ & 85688.4 &- & 310& 100&11 &b\\
H$^{13}$CN\,$(1-0)$ &	86340.2	& $0.38\pm 0.04$	&   300	& 100&3.54	& hf, t	\\
HCO\,$(1_{1,0}-0_{1,0})$&	86670.8& $0.32\pm0.03$	&  300& 100&2.99 & b, t\\
H$^{13}$CO$^+$\,$(1-0)$&	86754.3&$0.89\pm0.04$ &   300	& 100&8.39	 & b\\
SiO\,$(2-1)$	&	86847.0& $0.30\pm0.04$	&   300	& 100&2.87	 & b, t \\
C$_2$H\,$(1-0)$	&	87316.9& $20.61\pm 0.23$	&  $294.1\pm1.1$&$134.2\pm2.4$&  83.31 & hf \\ 
HNCO\,$(4_{0,4}-3_{0,3})$	&	87925.2&$0.26\pm0.45$&  300& 100& 2.41		& m, t	\\
HCN\,$(1-0)$	&	88631.8& $23.56\pm 0.40$&  $294.1\pm1.1$	& $113.6\pm2.6$&  182.00& hf	\\ 
HCO$^+$\,$(1-0)$ &	89188.6& $33.74\pm0.80$&  $292.8\pm1.6$& $128.8\pm3.7$&  236  	&	\\ 
HOC$^+\,(1-0)$	&	89487.4&$0.56\pm0.04$	  	&  300& 100	&  5.21		&	\\
HNC$(1-0)$	&	90663.6& $10.48\pm0.23$	&  $296.3\pm1.6$& $115.8\pm3.8$& 83.79	&	\\
HC$_3$N\,$(10-9)$	&  	90979.0&$1.42\pm0.03$&  $305.7\pm1.5$	& $109.5\pm3.4$	& 11.64&	\\
CH$_3$CN$\,(5_k-4_k)$&	91987.0&$0.24\pm0.72$& 300	&100&2.19& m, b	\\ 
H41$\alpha$ & 92034.4 &- & 280& 120&9 &b\\
N$_2$H$^+$\,$(1-0)$ &	93173.7& $1.65\pm0.04$	& $299.5\pm1.5$& 100& 15.35 & m  \\
CH$_3$CHO\,$(5_{1,5}-4_{1,4})$\,A &93580.9&$0.26\pm0.02$&300&100&2.12&m, t\\
C$^{34}$S\,$(2-1)$ &	96412.9& $0.58\pm0.03$	& $294.9\pm3.3$& 100	& 5.46	 	& 	\\
CH$_3$OH\,$(2_k-1_k)$&	96741.4&$1.01\pm0.03$& 300	& 100	& 9.17	 	& m	\\
CS\,$(2-1)$	&	97981.0& $9.97\pm0.16$&	$ 294.8\pm1.1 $	& 	$ 109.2\pm2.6 $	&84.66	& 	\\
H40$\alpha$ 	&99023.0	&$1.40\pm0.70$	&$302\pm8$	& 100	&13		& \\
SO\,$(2_1-3_2)$	&	99299.9& 	$1.11\pm0.11$	&  300	& 100	& 10.39	&	\\
HC$_3$N\,$(11-10)$	&  100076.4	& $1.42\pm0.03$	&$305.7\pm1.5$	& $109.5\pm3.4$& 12.26		& \\ 
CH$_3$CCH\,$(6_k-5_k) $ & 102548.0 & $3.48\pm0.11$&$314.6\pm1.9$ & 100& 32.01 &m \\	
H39$\alpha$	&106737.4	&$1.00\pm0.50$	&$307.2\pm5.4$	& 100	&9		& \\
CH$_3$OH\,$(0_0-1_{-1})$&108893.9&$0.254\pm0.009$& 300	& 100	& 2.39	 	& m	\\
HC$_3$N\,$(12-11)$	&  	109173.6	&$1.34\pm0.03$& $305.7\pm1.5$	& $109.5\pm3.4$		& 	12.14	& \\ 
C$^{18}$O\,$(1-0)$&	109782.2	& $5.78\pm0.10$	& $300.4\pm1.5$	& $103.0\pm3.5$& 52.35	&	\\ 
HNCO\,$(5_{0,5}-4_{0,4})$	&	109905.8	& $0.45\pm0.79$	&300	& 100	& 4.23	& m, t \\ 
$^{13}$CO\,$(1-0)$&	110201.4	& $28.25\pm0.39$	& $297.3\pm1.0$	& $107.1\pm2.3$& 240.09	&  \\ 
CH$_3$CN\,$(6_k-5_k)$ & 110383.5 & $0.28\pm0.85$ & 300&100 &2.56& m, t \\
C$^{17}$O\,$(1-0)$ & 112359.3&$0.38\pm0.07$ &300 & 100&3.55&t\\
CN\,$(1_{0,1}-0_{0,1})$	&113191.3	& $14.34\pm0.19$	& $291.5\pm1.1$	& $122.2\pm2.8$& 69.10	&m \\ 
CN\,$(1_{0,2}-0_{0,1})$	&113491.0	& $28.63\pm0.37$	& $291.5\pm1.1$	& $122.2\pm2.8$& 194.80	&m \\
NS\,$(3_{1,-1}-2_{1,1})$& 115153.9	&$0.73\pm0.13$	& 300	& 100&{\bf{6.11}}& m, b, t \\
$^{12}$CO\,$(1-0)$ &	115271.2	& $535.8\pm2.85$& $294.3\pm1.4$	& $94.4\pm3.1$& 3892.23 	& b	\\
H$\alpha$38 & 115274.4 &- & 300& 100&9 &b\\
NS\,$(3_{1,1}-2_{1,-1})$	& 115556.2	& $0.73\pm0.13$& 300&100& {\bf{5.55}}& m, t \\	
\hline
\end{tabular}

\begin{list}{}{}
\item[]  See caption of Table~\ref{TableM83} for details.
\end{list}{}{}

\label{GaussParM82}
\end{table*}

\begin{table*}[h!]
\caption{Results for {\bf{M\,51}} }

\centering
\begin{tabular}[!h]{lrrrrrrrrrrr} 
\hline
\hline
Line	&	Frequency	&	$I$ &    $V_{\rm LSR}$  	  & $\Delta V$  		& $T^{\rm peak}_{\rm MB}$ 		 &Comments	\\
	&	MHz		&	K\,km\,s$^{-1}$		&  km\,s$^{-1}$	  &   km\,s$^{-1}$	& 	 mK		&&		\\	
\hline
H$^{13}$CN\,$(1-0)$ &	86340.2	& $0.19\pm 0.04$	&  470& 130 	&1.38& hf, t	\\
C$_2$H\,$(1-0)$	&	87316.9&$1.26\pm0.07 $	&  $466.6\pm5.5$& $140.4\pm12.2$& 5.04	&hf \\ 
HCN\,$(1-0)$	&	88631.8&$5.39\pm0.10 $	&  $473.0\pm1.5$	& $129.1\pm3.6$&  	38.75	&hf	\\ 
HCO$^+$\,$(1-0)$ &	89188.6& $2.67\pm0.07$ 	&  $473.3\pm2.2$ 	& $132.0\pm5.1$&   18.98	& 	\\ 
HNC\,$(1-0)$	&	90663.6&$1.79\pm0.06$	&  $472.9\pm2.6$& $126.7\pm6.0$&  13.23		&	\\
N$_2$H$^+$\,$(1-0)$ &	93173.7&$0.74\pm0.04$	& $466.4\pm5.5$	& $146.8\pm13.0$& 4.70	& m  \\
C$^{34}$S\,$(2-1)$ &	96412.9&$0.17\pm0.04$	& 470 	& 130	& 	1.24	& t	\\
CH$_3$OH\,$(2_k-1_k)$&96741.4&$0.24\pm0.06$	& 470	& 130& 1.72& m, t	\\
CS\,$(2-1)$	&	97981.0&$1.04\pm0.05$	&	$ 464.8\pm 4.1$	& 	$125.6 \pm9.6 $	&7.74	&	\\
SO\,$(2_1-3_2)$	&	99299.9&$0.25\pm0.06$&  470	& 130	&1.80& t	\\
C$^{18}$O\,$(1-0)$&	109782.2	& $1.61\pm0.09$	& $474.3\pm4.7$	& $133.5\pm11.1$		&11.34&	\\ 
$^{13}$CO\,$(1-0)$&	110201.4	& $6.68\pm0.25$	& $469.7\pm3.0$	& $123.9\pm7.1$&50.40&  \\ 
C$^{17}$O\,$(1-0)$&	112359.3	& $0.30\pm0.06$ 	& 470	& 130		&2.14& t	\\ 
CN\,$(1_{0,1}-0_{0,1})$	&113191.3	& $2.55\pm0.04$& $468.4\pm1.3$	& $130.0\pm3.4$	 	&12.16& m\\ 
CN\,$(1_{0,2}-0_{0,1})$	&113491.0	& $5.12\pm0.07$& $468.4\pm1.3$	& $130.0\pm3.4$		&34.53& m\\
$^{12}$CO\,$(1-0)$ &	115271.2	& $50.14\pm1.53$	& $468.5\pm2.5$	& $120.0\pm5.7$& 379.30	& 	\\
\hline
\end{tabular}

\begin{list}{}{}
\item[] See caption of Table~\ref{TableM83} for details.
\end{list}{}{}

\label{GaussParM51}
\end{table*}

\begin{table*}[h!]
\caption{Results for {\bf{NGC\,1068}}}

\centering
\begin{tabular}[!h]{lrrrrrrrrrrrr} 
\hline
\hline
Line	&	Frequency	&	$I$ &  $V_{\rm LSR}$
& $\Delta V$ 		& $T^{\rm peak}_{\rm MB}$		& Comments	\\
	&	MHz		&	K\,km\,s$^{-1}$		&  km\,s$^{-1}$	  &   km\,s$^{-1}$	& mK&		&		\\	
\hline
H$^{13}$CN\,$(1-0)$ &	86340.2	& 	$0.78\pm 0.08$	&   $1088.4\pm11.2$	&  180	& 4.03  	& hf 	\\
HCO\,$(1_{1,0}-0_{1,0})$&	86670.8		& $0.49\pm0.12 $	&1100& 240		& 1.08	& t, b, hf\\
SiO\,$(2-1)$	&	86847.0& 	$0.60\pm0.07$	&  1100	&240	& 2.36		& t \\
HN$^{13}$C\,$(1-0)$ &	87090.8		& $0.19\pm0.05$	& 1100  	& 150	&  1.20 		&  t\\ 
C$_2$H\,$(1-0)$	&	87316.9& 	$7.80\pm0.10$&$1106.4\pm2.6$	& $239.8\pm5.3$		&  18.51 & b, hf \\ 
HNCO\,$(4_{0,4}-3_{0,3})$	&	87925.2& $0.57\pm0.89$&1100	&240		& 2.20		& m, t, b	\\
HCN\,$(1-0)$	&	88631.8& 	$23.55\pm 0.28$	&  $1112.8\pm1.6$	& $215.9\pm3.4$		&  77.84	& hf	\\ 
HCO$^+$\,$(1-0)$ &	89188.6		& 	$14.36\pm0.17$	&  $1113.7\pm1.9$ 	& $219.3\pm4.2$	&  56.30	&	\\ 
HOC$^+(1-0)$	&	89487.4& $0.31\pm0.06$  	&  1110	& $234.1\pm73.4$	&  1.23		&t	\\
HNC$(1-0)$	&	90663.6		& $7.78\pm0.10$	&  $1112.2\pm1.9$ 	& $225.1\pm4.4$		& 31.12	&	\\
HC$_3$N\,$(10-9)$	&  	90979.0&$0.71\pm0.03$	&  $1091.0\pm5.6$	& 190	& 3.52 &	\\
CH$_3$CN$\,(5_k-4_k)$&	91987.0& $0.59\pm0.42$	& 1100	&240	& 2.30		& m, t	\\ 
N$_2$H$^+$\,$(1-0)$ &	93173.7& 	$1.99\pm0.08$	& $1100.6\pm5.4$	& 240		& 7.62		& m  \\
C$^{34}$S\,$(2-1)$ &	96412.9		& $0.68\pm0.08$	& 1100 	& 240	& 2.64		& t	\\
CH$_3$OH\,$(2_k-1_k)$&96741.4&	$1.55\pm0.09$& 1100	& 240		& 5.93		& m	\\
CS\,$(2-1)$	&	97981.0&$3.95\pm0.09$& $1092.8\pm3.3$& 	$212.2\pm7.6$	& 17.14	& b	\\
SO\,$(2_1-3_2)$	&	99299.9& $0.47\pm0.06$	&  $1065.1\pm12.4$	& $139.7\pm29.0$	&  3.13		&	\\
HC$_3$N\,$(11-10)$	&  100076.4	& $0.80\pm0.04$	&  $1091.0\pm5.6$	& 190	& 3.95		& \\ 
HC$_3$N\,$(12-11)$	&  	109173.6	& 	$0.85\pm0.04$	&$1091.0\pm5.6$	&190		& 4.19		& t\\ 
C$^{18}$O\,$(1-0)$&	109782.2	& $	3.90\pm0.09$	& $1100.5\pm4.0$	& $258.0\pm9.4$		& 14.02		&	\\ 
$^{13}$CO\,$(1-0)$&	110201.4	& 	$13.10\pm0.24$	& $1100.8\pm2.9$	& $237.6\pm5.8$ 	& 49.36		&  \\ 
CH$_3$CN\,$(6_k-5_k)$&	110383.5	& $0.70\pm0.49$	& 1100	& 240		& 2.71		& m, t\\ 
CN\,$(1_{0,1}-0_{0,1})$	&113191.3	& 	$13.15\pm0.10$	& $1114.0\pm1.0$	& $224.4\pm2.5$	 	& 43.20		& m\\ 
CN\,$(1_{0,2}-0_{0,1})$	&113491.0	& $25.79\pm0.19$	&$1110.4\pm1.0$	& $224.4\pm2.5$	& 87.19	& m\\
$^{12}$CO\,$(1-0)$ &	115271.2	& $81.30\pm1.01$	& $1168.1\pm10.2$	&$122.5\pm10.8$		& 624.39		& m	\\
\hline
\end{tabular}

\begin{list}{}{}
\item[]  CO was fitted with a triple Gaussian, and its parameters refer
  to the central one. See caption of Table~\ref{TableM83} for more details. 
\end{list}{}{}

\label{TableNGC1068}
\end{table*}

\begin{table*}[h!]
\caption{Results for {\bf{NGC\,7469}} }

\centering
\begin{tabular}[!h]{lrrrrrrrrrrrr} 
\hline
\hline
Line	&	Frequency	&	$I$ &   $V_{\rm LSR}$ 	  & $\Delta V$  		&$T^{\rm peak}_{\rm MB}$	 & Comments	\\
	&	MHz		&	K\,km\,s$^{-1}$		&        km\,s$^{-1}$	  &   km\,s$^{-1}$	& mK &				\\	
\hline
C$_2$H\,$(1-0)$	&	87316.9		& $1.58\pm 0.14$	&  $4894.1\pm17.8$	& $233.0\pm35.1$&  4.19	& hf\\ 
HCN\,$(1-0)$	&	88631.8& 	$2.36\pm0.13$&$4854.3\pm8.7$	& $239.2\pm20.4$&  	9.18  &hf	\\ 
HCO$^+$\,$(1-0)$ &	89188.6		& $2.89\pm0.14$	&  $4847.3\pm8.0$ 	& $247.5\pm18.7$	&   10.90 &	\\ 
HNC\,$(1-0)$	&	90663.6&$1.08\pm0.12$	&  $4852.1\pm18.3 $ 	&  250& 4.07	& 	\\
CS\,$(2-1)$	&	97981.0		& 	$0.93\pm0.09$	&$ 4871.0\pm15.3 $	& 	$ 251.2\pm36.0 $	& 3.48 	& t	\\
C$^{18}$O\,$(1-0)$&	109782.2	& $0.42\pm0.08$	&4850	& 250		& 1.58		& t	\\ 
$^{13}$CO\,$(1-0)$&	110201.4	& $2.34\pm0.10$	& $4849.5\pm6.9$	& $245.9\pm16.1$ 	& 8.92		&  \\ 
CN\,$(1_{0,1}-0_{0,1})$	&113191.3	&$1.92\pm0.07$&  $4847.3\pm5.9$	& $248.0\pm14.4$		&6.33& m, t \\	
CN\,$(1_{0,2}-0_{0,1})$	&113491.0	& $3.85\pm0.49$& $4847.3\pm5.9$	& $248.0\pm14.4$	& 	14.14 &m \\
$^{12}$CO\,$(1-0)$ &	115271.2	& 	$58.12\pm1.51$	& $4847.0\pm7.6$	& $236.6\pm10.1$		&208.61& 	\\
\hline
\end{tabular}

\begin{list}{}{}
\item[]  See caption of Table~\ref{TableM83} for details.
\end{list}{}{}

\label{GaussParNGC7469}
\end{table*}

\begin{table*}[h!]
\caption{Results for {\bf{Arp\,220}}}

\centering
\begin{tabular}[!h]{lrrrrrrrrrr} 
\hline
\hline
Line	&	Frequency	&	$I$ &    $V_{\rm LSR}$ 	  &     $\Delta V$ 		& $T^{\rm peak}_{\rm MB}$ 		& Comments	\\
	&	MHz		&	K\,km\,s$^{-1}$		&  km\,s$^{-1}$	  &   km\,s$^{-1}$	& mK	&		\\	
\hline
H$^{13}$CN\,$(1-0)$ &	86340.2	& 	$1.65\pm 0.09$ &  $ 5364.1\pm11.1 $	& $328.0\pm25.6$ 	&4.46& hf 	\\
H$^{13}$CO$^+$\,$(1-0)$&	86754.3& $0.40\pm0.34$ & 5350&390	&0.95& b, t\\
SiO\,$(2-1)$	&	86847.0&$1.35\pm 0.18$	&  5350	& $363.8\pm44.0$	&3.42& b \\
C$_2$H\,$(1-0)$	&	87316.9& $3.84\pm 0.13 $&  $5317.3\pm11.5$	& $388.2\pm32.5$&  6.29	& hf \\ 
HCN\,$(1-0)$	&	88631.8& $11.06\pm 0.24$&  $5339.1\pm6.1$	& $448.1\pm13.2$&  18.02		& hf	\\ 
HCO$^+$\,$(1-0)$ &	89188.6& $4.62\pm0.21$  	&  $5312.8\pm14.3$ 	& 450		&   9.13	&	\\ 
HNC$(1-0)$	&	90663.6&  $7.78\pm0.34$  &  $5326.7\pm12.7$ 	& $378.8\pm27.8$& 	17.24	&	\\
HC$_3$N\,$(10-9)$	&  	90979.0&$3.38\pm0.09$&  $5331.4\pm7.4$	& $386.7\pm17.4$	& 8.11		&	\\
HC$_3$N,v7=1\,$(10_{-1}-9_{1})$	&  	91202.7&$0.39\pm1.01$&  5331& 390	& 1.67		&t	\\
HC$_3$N,v7=1\,$(10_{1}-9_{-1})$	&  	91333.4& $0.69\pm1.01$&  5331	& 390	& 1.67		&t	\\
CH$_3$CN$\,(5_k-4_k)$&91987.0&$1.41\pm0.20$	& 5350	&$386.1\pm83.2$&3.32& m	\\ 
N$_2$H$^+$\,$(1-0)$ &	93173.7& $2.36\pm0.42$	& $5353.0\pm8.9$	& $386.3\pm20.6$&5.42  & m  \\
C$^{34}$S\,$(2-1)$ &	96412.9& $1.19\pm0.14$	& $5393.6\pm35.7$	& 450	& 2.46& t	\\
CH$_3$OH\,$(2_k-1_k)$&	96741.4&	$0.88\pm0.09$	& 5350&390& 2.09& m, t	\\
CS\,$(2-1)$	&	97981.0&$4.15\pm0.07$	&$5342.6\pm4.7 $	&$ 408.1\pm10.7 $	&9.14	& 	\\
HC$_3$N\,$(11-10)$&  100076.4	& $4.17\pm0.11$	&  $5331.4\pm7.4$	& $386.7\pm17.4$		& 10.00		& \\ 
HC$_3$N,v7=1\,$(11_{1}-10_{-1})$	&  	100322.4&$0.99\pm1.01$&5331& 390	& 2.38		&	\\
HC$_3$N,v7=1\,$(11_{-1}-10_{1})$	&  	100466.2&$0.99\pm1.01$&5331& 390	& 2.39		&	\\
HC$_3$N\,$(12-11)$&109173.6	&$4.91\pm0.14$	& $5331.4\pm7.4$	& $386.7\pm17.4$		& 	11.78	& \\ 
HC$_3$N,v6=1\,$(12_{}-11_{})$	&109422.0& $1.36\pm1.01$&5331& 390	& 3.28		&	\\
HC$_3$N,v7=1\,$(12_{1}-11_{-1})$	&  	109598.8& $1.37\pm1.01$&5331& 390	& 3.29		&	\\
C$^{18}$O\,$(1-0)$&	109782.2	& 	$4.48\pm0.25$	& 5330	& 450&9.02&b	\\ 
HNCO\,$(5_{0,5}-4_{0,4})$	&	109905.8&$1.05\pm0.39$&$5344.2\pm84.4$	& $367.0\pm197.3$&2.64& m, b \\  
$^{13}$CO\,$(1-0)$&	110201.4	& 	$4.83\pm0.25$	& 5330	& 450 	& 9.71		&  b\\ 
CH$_3$CN\,$(6_k-5_k)$&110383.5	& 	$1.77\pm0.26$& 5350& $386.1\pm83.2$&4.18& m, b\\ 
CN\,$(1_{0,1}-0_{0,1})$&113191.3& 	$2.98\pm0.08$& $5365.4\pm10.4$& $532.4\pm19.8$& 	4.92	& m\\ 
CN\,$(1_{0,2}-0_{0,1})$&113491.0& 	$5.94\pm0.16$& $5365.4\pm10.4$	&$532.4\pm19.8$ & 	9.67	&m \\
$^{12}$CO\,$(1-0)$ &	115271.2	& 	$94.90\pm0.18$	& $5336.2\pm9.3$	& $429.0\pm0.9$	& 207.80	& 	\\
\hline
\end{tabular}

\begin{list}{}{}
\item[]  See caption of Table~\ref{TableM83} for details.
\end{list}{}{}

\label{GaussParArp220}
\end{table*}

\begin{table*}[h!]
\caption{Results for {\bf{Mrk\,231}} }

\centering
\begin{tabular}[!h]{lrrrrrrrrrrrrrrrr} 
\hline
\hline
Line	&	Frequency	&	$I$ &   $V_{\rm LSR}$  	  & $\Delta V$  		& $T^{\rm peak}_{\rm MB}$ 	 & Comments	\\
	&	MHz		&	K\,km\,s$^{-1}$		&        km\,s$^{-1}$	  &   km\,s$^{-1}$	& mK &	\\	
\hline
H$^{13}$CN\,$(1-0)$ &	86340.2	& $0.46\pm 0.10$	&  $12053.5 \pm29.4 $	&  210&2.02	& hf, t	\\
C$_2$H\,$(1-0)$	&	87316.9& $0.57\pm 0.11$	&  $12007.0\pm35.1$&210&  1.46 &hf, t \\ 
HCN\,$(1-0)$	&	88631.8& 	$2.06\pm 0.12$	&$11973.2\pm8.0$	& $207.8\pm18.3$&  8.86	 &hf	\\ 
HCO$^+$\,$(1-0)$ &	89188.6	&$1.34\pm0.09$ &$11951.0\pm10.9$	&  $234.0\pm25.1$	& 5.23	&   &	\\ 
HNC$(1-0)$	&	90663.6& $0.68\pm0.14$	&  $11975.6\pm28.3$	&  $204.7\pm65.9$& 3.06	& 	\\
HC$_3$N\,$(10-9)$	&  	90979.0&$0.25\pm0.06$&  $12067.4\pm33.4$	&210	& 1.10& t	\\
CS\,$(2-1)$	&	97981.0		&$0.44\pm0.09$ &$12048.0\pm27.4$	&210	& 1.96	& t	& 	\\
HC$_3$N\,$(11-10)$	&  100076.4	& $0.32\pm0.08$	&  $12067.4\pm33.4$	& 210& 1.40		& t\\ 
HC$_3$N\,$(12-11)$	&  109173.6	&$0.38\pm0.09$& $12067.4\pm33.4$	&	210	& 	1.68	&t \\ 
C$^{18}$O\,$(1-0)$&	109782.2	& $0.35\pm0.06$&$12056.6\pm23.7$	&210	& 1.57	& 		& 	\\ 
$^{13}$CO\,$(1-0)$&	110201.4	&$0.41\pm0.06$&$12086.1\pm21.2$	&210	& 1.81	& 		&  \\ 
CN\,$(1_{0,1}-0_{0,1})$	&113191.3& $0.31\pm0.21$&$12066.7\pm8.5$& $218.7\pm19.4$	& 1.32& m, t \\	
CN\,$(1_{0,2}-0_{0,1})$	&113491.0&$0.82\pm0.22$	& $12066.7\pm8.5$&$218.7\pm19.4$	& 3.46	& 	 m \\
$^{12}$CO\,$(1-0)$ &	115271.2	&$12.98\pm1.29$ &	$12038.6\pm7.8$	& 210	& 44.25	& 	\\
\hline
\end{tabular}

\begin{list}{}{}
\item[]  See caption of Table~\ref{TableM83} for details.

\end{list}{}{}

\label{GaussParMrk231}
\end{table*}

\end{appendix}

\end{document}